\documentclass[opre,nonblindrev]{informs3_mod}
\OneAndAHalfSpacedXI

\usepackage{endnotes}
\let\footnote=\endnote
\let\enotesize=\normalsize
\def\notesname{Endnotes}%
\def\makeenmark{\hbox to1.275em{\theenmark.\enskip\hss}}
\def\enoteformat{\rightskip0pt\leftskip0pt\parindent=1.275em
  \leavevmode\llap{\makeenmark}}

\usepackage{natbib}
 \bibpunct[, ]{(}{)}{,}{a}{}{,}%
 \def\bibfont{\small}%
 \def\bibsep{\smallskipamount}%
 \def\bibhang{24pt}%
\def\qed {{% set up
    \parfillskip=0pt % so \par doesnt push \square to left
    \widowpenalty=10000 % so we dont break the page before \square
    \displaywidowpenalty=10000 % ditto
    \finalhyphendemerits=0 % TeXbook exercise 14.32
    \leavevmode % \nobreak means lines not pages
    \unskip % remove previous space or glue
    \nobreak % don’t break lines
    \hfil % ragged right if we spill over
    \penalty50 % discouragement to do so
    \hskip.2em % ensure some space
    \null % anchor following \hfill
    \hfill % push \square to right
    $\Halmos$% % the end-of-proof mark
    \par}} % build paragraph

\usepackage[bf,sf]{caption}
\usepackage{subcaption}

\newcommand{\E}{\mathbb{E}}
\renewcommand{\P}{\mathbb{P}}

\newcommand{\eqD}{\overset{d}{=}}

\newcommand{\bbf}{\mathbf{b}}
\newcommand{\pbf}{\mathbf{p}}

\newcommand{\Ind}{\mathbf{1}}

\newcommand{\eps}{\epsilon}
\newcommand{\set}[1]{\left\{#1\right\}}
\newcommand{\abs}[1]{\left|#1\right|}
\newcommand{\goesto}{\rightarrow}

\newcommand{\Nbb}{\mathbb{N}}
\newcommand{\Rbb}{\mathbb{R}}

\newcommand{\Acal}{\mathcal{A}}
\newcommand{\Jcal}{\mathcal{J}}
\newcommand{\Ncal}{\mathcal{N}}
\newcommand{\Scal}{\mathcal{S}}

\newcommand{\erf}{\text{erf}}

\TheoremsNumberedThrough     % Preferred (Theorem 1, Lemma 1, Theorem 2)
\EquationsNumberedThrough    % Default: (1), (2), ...
\ECRepeatTheorems

\begin{document}
%%%%%%%%%%%%%%%%

%\setcounter{page}{1}
%\DOI{10.1287/xxxx.0000.0000}%

\RUNAUTHOR{Master, Chan, and Bambos}

\RUNTITLE{Myopic Scheduling of Jobs with Decaying Value}

\TITLE{Myopic Policies for Non-Preemptive Scheduling of Jobs with
  Decaying Value}

\ARTICLEAUTHORS{%
  \AUTHOR{Neal Master}
  \AFF{Department of Electrical Engineering, Stanford University, \EMAIL{nmaster@stanford.edu}}
  \AUTHOR{Carri W. Chan}
  \AFF{Decision, Risk, and Operations, Columbia Business School, \EMAIL{cwchan@columbia.edu}}
  \AUTHOR{Nicholas Bambos}
  \AFF{Department of Management Sciences \& Engineering and Department of Electrical Engineering, Stanford University, \EMAIL{bambos@stanford.edu}}
}

\ABSTRACT{%
  In many scheduling applications, minimizing delays is of high
  importance. One adverse effect of such delays is that the reward for
  completion of a job may decay over time.  Indeed in healthcare
  settings, delays in access to care can result in worse outcomes,
  such as an increase in mortality risk. Motivated by managing
  hospital operations in disaster scenarios, as well as other
  applications in perishable inventory control and information
  services, we consider non-preemptive scheduling of jobs whose
  internal value decays over time.  Because solving for the optimal
  scheduling policy is computationally intractable, we focus our
  attention on the performance of three intuitive heuristics: (1) a
  policy which maximizes the expected immediate reward, (2) a policy
  which maximizes the expected immediate reward rate, and (3) a policy
  which prioritizes jobs with imminent deadlines. We provide
  performance guarantees for all three policies and show that many of
  these performance bounds are tight. In addition, we provide
  numerical experiments and simulations to compare how the policies
  perform in a variety of scenarios. Our theoretical and numerical
  results allow us to establish rules-of-thumb for applying these
  heuristics in a variety of situations, including patient scheduling
  scenarios.
}

\maketitle

\clearpage
\section{Introduction}\label{sec:intro}
%\subsection{Motivation and Summary}
Managing delays in queueing networks has been the focus of a large
body of work (e.g.  \cite{Mandelbaum_cmu_2004},
\cite{Dewan_delay_1990}, and \cite{VanMieghem_Capcity_2003} among many
others).  Typically, the undesirability of and dissatisfaction due to
incurred delays serve as the primary motivation for minimizing delays.
However, there are other adverse effects of delays, which often are
not accounted for.  For instance, in healthcare settings, delays in
access to care can result in deterioration of a patient's health
state, thereby reducing the efficacy of the resulting care. In this
work, we consider how to prioritize jobs (e.g. patients) when the
reward for completing any particular job decreases over time.

Delays in healthcare are rampant. A study by
\cite{Poon_TestResults_2004} indicates that over 60\% of physicians
reported dissatisfaction in the timeliness of test results, which can
create treatment delays and, ultimately, lead to increased patient
mortality and increased healthcare costs.  Indeed, in the case of
intensive care, delays in treatment often lead to deterioration of
patient health and this can eventually reduce the efficacy of various
treatments \citep{McQuillan_Inquiry_1998}. This also occurs for
cardiac arrest \citep{Chan_Defib_2008, Buist_CardiacArrest_2002},
angioplasty for acute myocardial infarctions
\citep{Luca_Infarction_2004}, and is particularly true for children
\citep{Sharek_Children_2007}. In this work, we capture the impact of
delayed treatment on mortality risk and other health outcomes by
allowing for the reward for completing a job to decay arbitrarily. In
contrast to our work here, recent work by \cite{chan_delay_15}
examines the impact of delayed treatment on service time.

For a specific scenario in healthcare where scheduling of jobs with
decaying values is of interest, we start by considering patient triage in the
aftermath of mass casualty incidents. In these situations, medical
resources are overwhelmed by a sharp increase in demand. In both
civilian and military situations, medical personnel, operating rooms,
and ambulances need to be judiciously allocated so as to minimize the
number of deaths and permanent injuries. Treatment delays will reduce
survivial probabilities so rapid scheduling is a necessity. Triage
practices have evolved over time, but must continue to advance as new
disasters and new technologies can render previous strategies obsolete
\citep{Iserson_TriageI_2007, Moskop_TriageII_2007}.  There has been
recent work examining patient triage in disaster scenarios by the
operations management community
(e.g. \cite{Argon_Triage_2008,Argon_EORMS_2011,Chan_Burns_2013}); yet,
none have considered an arbitrary decay in reward as we do here.

While the primary motivation for our work is patient triage in mass
casualty events, we note that the reward decay dynamics we consider in
this work extend to other applications as well.  For example, in food
processing, perishable food items decay in market value as their
expiration dates approach and efficient scheduling is necessary to
maximize profits.  Managers must account for variations in customer
orders, equipment availability, raw materials, deliveries, processing
rates, and food freshness when making decisions regarding the
production of perishable foods like ice cream and yogurt
\citep{Jakeman_Dairy_1994}.  As another example, scheduling jobs with
deadlines has been of particular interest in information services and
computing. In this case, the value of a job decays according to a step
function. For example, deadlines are useful for ensuring high level of
quality for customers who are streaming multimedia
\citep{Dua_CAEDD_2007, Dua_CD2_2010}.  Deadlines have also been
considered in more general ``data broadcast'' problems which lead to a
number of combinatorial optimization problems
\citep{Kim_Broadcast_2004, Zheng_Broadcast_2006}.

Though the aforementioned applications are quite varied, they share a
number of key similarities. In each case, we need to dynamically
schedule jobs with decaying value for processing/service. The value of
the jobs decays over time and we seek to capture as much of this value
as possible. Motivated primarily by patient scheduling in disaster
scenarios, we choose to focus on non-preemptive scheduling of a
``clearing system'' in which all jobs are present at the initial time
(see \cite{Argon_EORMS_2011} and the references therein). To the best
of our knowledge, we are the first to account for the fact that jobs
each have an internal value which may decay over time. For each job,
the function which governs this decay is deterministic and
non-increasing; however, the manner of this decay is permitted to be
\emph{arbitrary}. 

The arbitrary decay generalizes the idea of having jobs with deadlines
which is a common modeling construct
(e.g. \cite{Argon_Triage_2008,Chan_Burns_2013,Dua_CAEDD_2007,Kim_Broadcast_2004}).
When jobs have deadlines, this corresponds to step-wise value decay:
the internal value of the job will abruptly transition from full value
to zero value after the deadline. This sharp transition can be thought
of as a ``hard'' deadline. In constrast, our model allows for ``soft''
deadlines. That is, rather than having the internal value of a job
\emph{abruptly} decay from full value to zero value, the decay
functions in our paper allow for the internal value of a job to
\emph{gradually} decay from full value to zero value. The time at
which the job reaches zero value can still be thought of as a
deadline, but because the transition from full value to zero value is
gradual, the deadline is ``soft'' rather than ``hard.'' The arbitrary
decay associated with soft deadlines allows for additional modeling
flexibility beyond what is allowed by hard deadlines. Our model allows
for both soft and hard deadlines as well as heterogeneity amongst the
jobs in the system, thus offering a substantial generalization over
previously studied scheduling models. Soft deadlines have recently
been considered in some service rate control problems (e.g.
\cite{Master_2014ICC, Master_2015ACC}) but not in scheduling problems.

\begin{figure}
  \FIGURE%
  {\includegraphics*[width=0.5\textwidth]{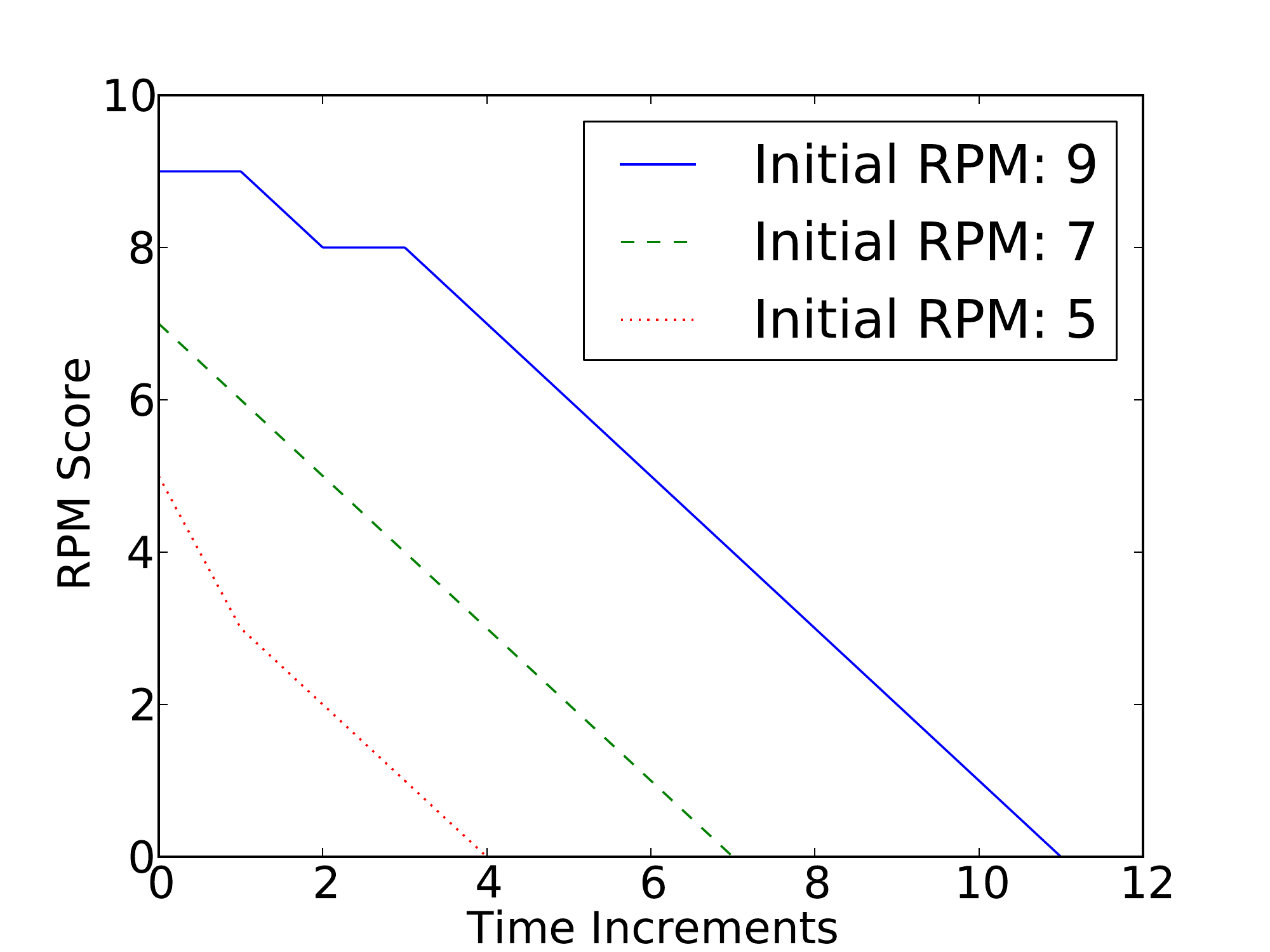}}%
  {Example RPM (Respiratory rate, Pulse rate, and Motor response)
    curves from Table 5 of
    \cite{Sacco_Triage_2005}\label{fig:sacco}. }%
  {The RPM scores provide a metric for patient health decay. The
    scores take integer values $\set{0, 1, \hdots, 12}$ and decay over
    30 minute time increments.}
\end{figure}

This modeling generalization is particularly important in patient
scheduling applications. In the patient scheduling literature
(e.g. \cite{Argon_Triage_2008,Chan_Burns_2013}), hard deadlines are
used to model the time of mortality due to the injury/ailment at
hand. However, this may not capture all of the nuances associated with
the patient health in disaster scenarios, and soft deadlines may be
more appropriate. For example, consider the patient triage scheme
developed by \cite{Sacco_Triage_2005}. By consulting a group of
physicians, \citeauthor{Sacco_Triage_2005} design a ``health score''
which they call RPM (Respiratory rate, Pulse rate, and Motor response)
which decays over 30 minute time intervals. In Figure~\ref{fig:sacco}
we have plotted a few of the decaying RPM curves from \cite[Table
5]{Sacco_Triage_2005}. Note that the decay is sometimes linear but not
necessarily. This motivates the arbitrary decay in our model. We
emphasize that while \citeauthor{Sacco_Triage_2005} prioritize
patients based on their gradually decaying health, they do so in a
static manner and do not allow for dynamic patient scheduling. A key
feature of our model is that we incorporate decaying job value and
dynamic non-preemptive scheduling. We will discuss
\cite{Sacco_Triage_2005} more in the literature review.

% For example, in a study by \cite{Luca_Infarction_2004} on acute
% myocardial infarctions treated primarily by angioplasty, a
% statistically significant correlation was found linking treatment
% delays and elevated 1-year mortality risk. As a result, the long-term
% value of treatment decays gradually over time rather than abruptly and
% the decay is not related to immediate risk of mortality. The soft
% deadlines in our model capture this kind of value decay while hard
% deadlines (which were the focus of previous models) do not.

While soft and hard deadlines are useful modeling techniques
(particularly for patient scheduling applications), we will show that
maximizing the total value over time is computationally intractable.
As such, we turn our attention to a number of intuitive, yet
sub-optimal scheduling heuristics. In doing so, we wish to examine how
well one can expect heuristics which do not account for future system
dynamics to perform. Additionally, we aim to identify which heuristics
are most effective for various different situations. More
specifically, we consider three different heuristics:
\begin{enumerate}
\item Whenever there is a free server, the {\bf greedy policy}
  schedules the job which maximizes the expected reward generated by
  the completion of that job, where the expectation is taken with
  respect to the job's service time distribution.  As such, the reward
  considered by this heuristic accounts for the decay in value of the
  job.

\item Whenever there is a free server, the {\bf rate greedy policy}
  schedules the job with the maximum expected reward rate. This simply
  takes the expected reward generated by the completion of the job as
  considered by the greedy policy and divides it by the expected
  service time of the job in order to estimate the reward generated
  \emph{per unit of time} during the processing of the scheduled job.

\item Whenever there is a free server, the {\bf Earliest Deadline
    First (EDF) policy}, schedules the job whose value decays to 0
  soonest.  In order for this policy to be well defined, we must
  assume that for each job there is a finite time at which the value
  generated for completing the job is equal to 0, i.e. each job has a
  final deadline. The EDF policy schedules jobs whose deadlines are
  most imminent and have not yet passed.
\end{enumerate}

We provide performance guarantees for each heuristic and are able to
show that in some situations, these bounds are tight.  We further
explore the performance across the heuristic policies through
simulation and provide some rules-of-thumb for when each of the
policies is most appropriate. In particular, our simulations support
the following rules:
\begin{itemize}
\item The rate greedy policy is the most robust heuristic in that it
  seems to always performs well and that for large numbers of jobs
  with high levels of heterogeneity, it will perform better than the
  other two heuristics.

\item The greedy policy is also a very good heuristic. For many small
  scale problems, it can outperform the other heuristics. While the
  rate greedy policy generates more reward for large problems, the
  performance of the greedy policy is not far behind.

\item EDF can outperform the greedy policies, but it is not very
  robust. We identify a few scenarios in which the performance of EDF
  is on par with the two greedy policies, but we note that slight
  deviations from these scenarios will lead to poor performance for
  EDF.
\end{itemize}

The remainder of the paper is outlined as follows.  We conclude this
section with a brief overview of some related literature. In
Section~\ref{sec:model} we introduce our model and the scheduling
problem we consider. We show that while the problem is well-posed, it
is computationally intractable. As such, we turn to heuristic policies
in Section~\ref{sec:heuristics}. We provide performance guarantees for
three different heuristics and briefly discuss how the proofs can be
leveraged to provide performance guarantees for other myopic
policies. In addition to the general performance guarantees, we show
that there are situations in which each policy is better than the
other two. We explore this more in Section~\ref{sec:peva} via
numerical examples. We use these results to extract rules-of-thumb for
understanding when each heuristic should (or should not) be used. We
conclude in Section~\ref{sec:conclusion}. Proofs of our mathematical
results are given in the appendix.

\subsection{Literature Review}\label{ssec:lit}
Our work is related to the healthcare operations management literature
on patient triage and scheduling.  While our primary motivation is
scheduling patients in mass casualty events, our model also has some
similarities to work done in perishable inventory control and
information services.  More generally, our work is related to
theoretical work in scheduling and the evaluation of heuristic
policies.

In the healthcare operations management literature, clearing models
for jobs with value decay have been used to study triage and patient
scheduling in mass casualty incidents. \cite{Argon_Triage_2008}
consider a clearing system where jobs are characterized by random
service times as well as random deadlines. They show that if the jobs
can be ordered in a particular way such that the job with the shortest
deadline also has the shortest service time, then an optimal policy is
to give priority to the most ``time-critical'' job. Unfortunately,
jobs do not always exhibit this ordering and, in general, the
time-critical first (TCF) heuristic performs poorly. Our model of job
decay is quite different as we allow for arbitrary, deterministic,
non-increasing functions rather than binary functions which are
stochastic. We demonstrate that like the TCF policy, the EDF policy
also performs poorly in general, but it can also do well in certain
special cases.  Moreover, we also consider the performance of other
policies and provide performance guarantees for them.

In related work, \cite{Mills_START_2013} consider a fluid model for
patient triage which considers dynamic patient survival
probabilities. Their model focuses on a finite number of patient
classes each with time varying rewards for service completion. By
controlling the service rates for each class, they seek to maximize
the long-term reward.  We also consider dynamic patient scheduling.
However, instead of using fluid models, which necessarily only capture
``average'' behavior (see \cite{Gamarnik_Fluid_2010} for a survey), we
consider a stochastic model and evaluate the performance of a number
of heuristics. Additionally, while \citeauthor{Mills_START_2013} focus
on ambulance transportation, we have calibrated our simulations to
provide insight into scheduling surgical procedures in mass casualty
events (see Section~\ref{ssec:patients}).

Similar to our model, \cite{Sacco_Triage_2005} consider how to do mass
casualty triage given patients have a ``health score'' which decays
over time.  This score decreases in a deterministic fashion over the
finite time horizon and maps to a survival probability. The
deterministic health score system is determined by the Delphi method,
an iterative survey technique which is often used to aggregate expert
opinions in a quantitative manner (see \cite{Linstone_Delphi_1975} for
an overview of the Delphi method). This suggests that precise formulae
for the deterministic value decay in our model could be determined in
a similar fashion.  Given various capacity constraints, a linear
program is solved to decide how many patients of each score should be
scheduled in each time slot so as to maximize the expected number of
survivors. Note that this linear program is solved once at the
beginning of the time horizon. As a result, while this technique does
allow for arbitrary patient health decay, it does not allow for
dynamic scheduling decisions. In contrast, the policies we consider,
albeit myopic, are dynamic and can adapt to a changing environment.

Another line of research in patient scheduling has been to rely on
sophisticated computational techniques to approximately compute
optimal policies. For example, \cite{Patrick_MDP_2008} pose a patient
scheduling problem as a Markov Decision Problem (MDP) and use linear
programming based Approximate Dynamic Programming (ADP) techniques to
find high performance policies. We also leverage the theory of MDPs
for studying our scheduling model. However, rather than focus on
computational techniques, we investigate the efficacy of simple
heuristics which are easy to implement.

While our primary motivation is healthcare operations, we note that
our model also captures some features present in other types of
systems. In inventory control, \cite{Federgruen_Inventory_2015}
recently showed that ``shelf-age-dependent'' and ``delay-dependent''
cost structures are equivalent to traditional ``level dependent'' cost
structures. These time-varying costs are conceptually similar to the
value decay functions which we employ, but their focus is on inventory
models rather than patient scheduling systems. As another example, we
note that jobs with deadlines have played an important role in the
information technology literature.  For example, \cite{Dua_CAEDD_2007}
consider the problem of scheduling multiple traffic streams over a
wireless downlink. Their solution can be thought of as an algorithmic
incarnation of our rate greedy heuristic which accounts for
time-varying parameters and user preferences.  Our work expands on
these ideas significantly by considering more general value decay
functions and providing performance guarantees.

In the healthcare context, it is typical to consider non-preemptive
scheduling and specifically in mass casualty incidents, it is typical
to consider clearing systems (see \cite{Argon_EORMS_2011} and the
references therein). The key idea is that after a disaster, the
victims will undergo triage in one large batch and once a patient is
undergoing a procedure it is unsafe to preempt service. If the
scheduling were preemptive, then we could apply the theory of
stochastic depletion problems \citep{Chan_MathOR_2009}. However,
because the scheduling discipline is non-preemptive, when a server
begins work on a job, it will continue until the job is complete.  In
this sense, scheduling decisions tend to have a greater impact on the
future evolution of the system and intuitively, this makes
non-preemptive scheduling a more difficult problem. In particular, we
will later see that computing an optimal non-preemptive scheduling
policy for our system is at least as difficult as solving a broad
class of NP-hard combinatorial optimization problems.

In the more traditional case where the value of each job does not
decay over time and there is a single server, job scheduling problems
are often cast in the framework of multi-armed bandit problems in
which the Gittens Index Theorem allows for efficient computation of
optimal solutions (see \cite{Gittins_MAB_2011} and the references
therein). When the internal state of each job evolves over time, we
are faced with a restless bandit problem
\citep{Whittle_RestlessBandits_1988}. Index policies for restless
bandit problems have been shown to be asymptotically optimal (for
large numbers of jobs) in the case that the ratio between of the
number of jobs and servers is constant and a certain differential
equation describing a fluid approximation of the system is globally
stable \citep{Weber_Index_1990}. However, verifying the stability of
this differential equation is not always straightforward. In addition,
the indices in this result come from the Lagrange multipliers of an
associated optimization problem and as a result, this type of policy
may not be easy to implement in applications.

Still, there exists special cases where restless bandit problems admit
solutions which are easy to compute and implement. In a Markovian
setting where the reward for serving a user decays exponentially as a
function of the user's sojourn time, a greedy policy which seeks to
maximize immediate expected rewards is an optimal policy
\citep{Dalal_Impatient_2005}. This type of greedy policy can also be
framed as a $c\mu$-type policy (see \cite{Walrand} for a review of
such policies). The $c\mu$-type policy is shown to be optimal in a
heavy traffic setting \citep{Mandelbaum_cmu_2004}. These results
partially motivate the work in this paper. We consider similar greedy
policies but with arbitrary job value decay. In this more general
setting, the greedy policies are not optimal and so we turn out
attention to establishing bounds on their sub-optimality.

\section{Model Formulation}\label{sec:model}
We now formally introduce our model and discuss the optimization of
the scheduling problem we consider here.  While we are primarily
motivated by patient triage in mass casualty events, we present a
general model and connect back to the healthcare setting using
simulation in Section \ref{ssec:patients}.

\subsection{System Dynamics and Dynamic Programming Formulation\label{ssec:dp}}
We consider a set of $J$ jobs indexed by $j \in \Jcal = \set{1, 2,
  \hdots, J}$. Job $j \in \Jcal$ has a random service time $\sigma_j$
taking values in $\set{1, 2, 3, \hdots}$. Let the distribution of
$\sigma_j$ be $F_j(\cdot)$. The processing times are statistically
independent and each has a finite mean. There are $N$ identical
processors/servers indexed by $n \in \Ncal = \set{1, 2, \hdots,
  N}$. Each processor has unit service rate and can process a single
job at a time. Service is non-preemptive in the sense that once a
processor begins work on a job, the processor will continue to work on
this job until the job is completed. Time is slotted and indexed by $t
\in \set{0, 1, 2, \hdots}$.

Let $B_j(t)$ be the residual service time of job $j$ at time $t$ and
let $B(t) = (B_1(t), \hdots, B_J(t))$ be the total system backlog. We
know that $B(0) = (\sigma_1, \hdots, \sigma_J)$ and if all jobs
have completed at time $T$, $B(T) = (0, \hdots, 0)$. Each job service
time is random and the realization can only be seen after the job has
completed processing. Therefore, to make optimal scheduling decisions,
we must track the \emph{observable backlog vector} denoted by
$b(\cdot)$:
\begin{equation}
  \begin{split}
    b_j(t)= \left\{
      \begin{array}{cl}
        \bot, &\mbox{job $j$ hasn't begun processing}\\
        t', &\mbox{job $j$ began processing at time $t' < t$ and is still being processed}\\
        \top, &\mbox{job $j$ has completed processing}
      \end{array}
    \right.
  \end{split}
\end{equation}
Because the service time distributions are known, $b(t)$ encodes the
information necessary to determine the distribution of $B(t)$. We next
define the state $p_n(t)$ of processor $n \in \Ncal$ as follows:
\begin{equation}
  p_n(t)= \left\{
    \begin{array}{ll}
      j, & \mbox{if processor $n$ is executing job $j\in\cal J$ at the beginning of time slot $t$} \\
      0, & \mbox{if processor $n$ is free at the beginning of time slot $t$}
    \end{array}
\right.
\end{equation}
The \emph{processor state vector} is then defined as $p(t) = (p_1(t),
\hdots, p_N(t))$. With these definitions, we can take the system state
at time $t$ as
\begin{equation}
  s_t = (b(t), p(t))
\end{equation}
where we will sometimes simply write $s$ when the dependence on $t$ is
understood. We denote the state space
\begin{equation}
  \Scal
  \subseteq (\set{\bot} \cup \Nbb \cup \set{\top})^J \times \set{0, 1, \hdots, J}^N.
\end{equation}
Note that at the beginning of each time slot, we can schedule job $j$
on processor $n$ if and only if $b_j(t) = \bot$ and $p_n(t) =
0$. %Since $b(t)$ also encodes the distribution of $B(t)$, $s_t \in
%\Scal$ is a sufficient description of the stochastic evolution of the
%system which can be used to dynamically make scheduling decisions.

Given state $s$, let $\Acal(s)$ be the set of feasible scheduling
actions. A scheduling action is a set of pairings between jobs and
processors so if $A \in \Acal(s)$, then $(j, n) \in A$ only if $b_j(t)
= \bot$ and $p_n(t) = 0$.  In addition, since each processor can only
work on a single job at a time, if $(j, n) \in A$ and $(j', n') \in A$
then $j \ne j'$ and $n \ne n'$. If the system is in state $s$ at time
$t$ and action $A$ is take, let $S^+(s, A)$ be the random state at
time $t+1$.

If job $j$ is completed at the end of time slot $t$, we garner a
non-negative reward $v_j(t)$. We assume $v_j(\cdot) \geq 0$ and that
$v_j(\cdot)$ is non-increasing. Therefore, $v_j(\cdot)$ defines the
deterministic time-varying value of job $j$. For example, if job $j$
has a value $\nu_j$ and a deterministic deadline $d_j$, then we can
take $v_j(t) = \nu_j \Ind_{\set{t \leq d_j}}$.

Recall that if job $j$ is scheduled on processor $n$ at the beginning
of time slot $t$, it will complete processing at the end of time slot
$t + \sigma_j$. Therefore, if the scheduler chooses action $A \in
\Acal(s)$, the resulting reward will be
\begin{equation}
  R_t(s, A) = \sum_{(j, n) \in A}v_j(t + \sigma_j).
\end{equation}
We restrict our attention to deterministic policies $\pi \in \Pi$ such
that if $s \in \Scal$, then $s \mapsto \pi_t(s) \in \Acal(s)$. We can
now define the value function of policy $\pi$ as
\begin{equation}
  \label{eq:value}
  V_t^\pi(s)
  = \E\left[\left.
      \sum_{\tau = t}^\infty R_\tau(s_\tau, \pi_\tau(s_\tau))
    \right|s_t = s\right].
\end{equation}
The optimal value function is then defined as
\begin{equation}
  \label{eq:opt_value}
  V_t^*(s) = \max_{\pi \in \Pi} V_t^\pi(s).
\end{equation}
Any optimal policy which achieves this supremum is denoted $\pi^*$.

\subsection{Preliminary Mathematical Results\label{ssec:prelim}}
Our problem formulation lends itself to an MDP approach. Given the
recursive optimality equations (i.e. the Bellman equation) associated
with an MDP, one can compute the optimal value function and find an
optimal policy via standard techniques (value iteration, policy
iteration, and linear programming \citep{Bertsekas_VolII}). Our first
theorem characterizes the value function in terms of a Bellman
equation, thus demonstrating that this is (in principle) a valid
approach to the problem.
\begin{theorem}
  \label{thrm:well-posed}
  The quantities in \eqref{eq:value} and \eqref{eq:opt_value} are
  well-defined. An optimal policy $\pi^*$ exists and is characterized
  by the following Bellman equation:
  \begin{equation}
    \label{eq:bellman}
    V_t^*(s) = \max_{A \in \Acal(s)}\set{\E[R_t(s, A)] + \E[V_{t+1}^*(S^+(s, A))]}.
  \end{equation}
\end{theorem}

Our next theorem shows that this approach is not computationally
tractable. Dynamic programming problems with large state and action
spaces typically suffer from the \emph{curse of
  dimensionality}. However, our particular problem is possibly even
more difficult to solve. As mentioned before, requiring non-preemptive
scheduling adds a combinatorial ``twist'' to the problem. A broad
class of ``knapsack'' problems (see \cite{Martello_Knapsack}) can be
reduced to our problem and as a result, our general scheduling problem
is NP-hard (see \cite{CLRS} for an introduction to complexity theory
and NP-hardness).
\begin{theorem}
  \label{thrm:NP-hard}
  Computing $\pi^*$ is NP-hard.
\end{theorem}

\section{Heuristic Policies and Performance Guarantees}\label{sec:heuristics}
We have just seen that while the non-preemptive scheduling problem can
be solved in principle, it is unlikely that there is a computationally
tractable way of doing so. As such, we turn our attention to
heuristics which are intuitive and easy to use in practice. Unlike the
optimal policy, which solves \eqref{eq:bellman}, the heuristics we
consider ignore the impact of the scheduling decision on the future
value.  We focus on these particular heuristics because they have been
studied in other contexts (e.g. \cite{Dua_CD2_2010,
  Dalal_Impatient_2005, Mandelbaum_cmu_2004}) but we will briefly
comment on how our proofs can be extended to other policies. The
heuristics are defined as follows:
\begin{enumerate}
\item A {\bf greedy policy} selects the jobs with the largest expected reward:
  $$
  \pi_t^G(s) \in \argmax_{A \in \Acal(s)}
  \set{\E[R_t(s, A)]}
  =
  \argmax_{A \in \Acal(s)}
  \set{\sum_{(j, n) \in A} \E[v_j(t + \sigma_j)]}
  $$
\item A {\bf rate greedy policy} selects the jobs with the largest expected reward rate:
  $$
  \pi_t^g(s) \in \argmax_{A \in \Acal(s)}
  \set{\sum_{(j, n) \in A} \frac{\E[v_j(t + \sigma_j)]}{\E[\sigma_j]}}
  $$
\item An {\bf Earliest Deadline First (EDF) policy} schedules the jobs with the most imminent deadline:
  $$
  \pi_t^{EDF}(s) \in \argmin_{A \in \Acal(s)} \set{
    \sum_{(j,n) \in A}\frac{d_j - t}{\Ind_{\set{d_j \geq t}}}
  }
  $$
  where $d_j = \arg\sup_t \{v_j(t) > 0\} < \infty$ and we consider
  $c/0 = \infty$ for any $c \in \Rbb$. Note that this policy is only
  well-defined when the reward functions have finite deadlines. That
  is, for each job $j$, there is a finite integer $d_j$ such that
  $v_j(d_j) > 0$ but $v_j(d_j + 1) = 0$. Under the EDF policy, if a
  deadline has passed, it will not schedule this job until after all
  other jobs have been scheduled.
\end{enumerate}

For each policy, we assume that ties are broken in an arbitrary
fashion (e.g. assigning an ordering to actions and taking the
``smallest'').  While each of these strategies is intuitive, they can
lead to drastically different performance. For example, consider when
the service requirements are identically distributed with distribution
$F(\cdot)$ and $v_j(t) = \Ind_{\set{t \leq d_j}}$ for given deadlines
$\set{d_j}_{j \in \Jcal}$. Because the service times are IID,
$\pi^G$ and $\pi^g$ coincide with:
$$
\pi_t^G(s) = \pi_t^g(s) =
\argmax_{A \in \Acal(s)}\set{\sum_{(j, n) \in A} \E[v_j(t + \sigma_j)]}
= \argmax_{A \in \Acal(s)}\set{\sum_{(j, n) \in A} F(d_j - t)}.
$$
Because $F(\cdot)$ is monotonically increasing, $\pi^G$ and $\pi^g$
will schedule jobs with the \emph{latest} deadlines, the idea being
that these jobs are most likely to complete processing before their
deadline and generate reward. In this sense, $\pi^{EDF}$ is the
opposite of $\pi^G$ and $\pi^g$. While each strategy is intuitive in
its own way, it isn't immediately clear which policy will have better
performance under different situations. The following examples
demonstrate that no one of the heuristics dominates any of the others
-- there are situations in which each heuristic has a greater expected
value.

\begin{example}[$\pi^G$ can be better than $\pi^g$]
  \label{ex:pi_G_v_pi_g}
  Consider a system at $t = 0$ with $J = 2$ and $N = 1$. Let job 1 be
  characterized by
  $$
  \sigma_1 = \left\{
    \begin{array}{rl}
      1, &\hbox{w.p. } 0.99\\
      100, &\hbox{w.p. } 0.01
    \end{array}
  \right.,\quad
  v_1(t) = \Ind_{\set{t \leq 1}}
  $$
  and job 2 be characterized by
  $$
  \sigma_2 = 1, \quad
  v_2(t) = (1 - \delta)\Ind_{\set{t \leq 1}}
  $$
  for any $\delta \in (0.01, 0.5)$.  A simple computation shows that
  $\pi^G$ will schedule job 1 and then job 2 so that $V_0^G(s) = 0.99$
  while $\pi^g$ will schedule job 2 then job 1 so that $V_0^g(s) = 1 -
  \delta$. Therefore, $V_0^G(s) > V_0^g(s)$.
\end{example}

The intuition behind $\pi^g$ is that because of the decaying value of
each job, one should try to maximize the immediate reward per unit
time rather than merely the immediate reward. To this end, $\pi^g$
maximizes $\E[v_j(t + \sigma_j)]/\E[\sigma_j]$ rather than $\E[v_j(t +
\sigma_j)]$. Example~\ref{ex:pi_G_v_pi_g} shows how this estimate of
the reward rate can go wrong. When $\sigma_j$ has high variance, the
expected reward divided by the expected service time is not a good
estimate of the reward rate. Indeed, in the previous example the
standard deviation of $\sigma_1$ was nearly $10$. This issue is
exacerbated by the fact that the time horizon is short (effectively
one time slot) relative to the standard deviation. Since $\pi^G$
maximizes the immediate reward rather than the
immediate reward rate, $\pi^G$ is able to outperform $\pi^g$. The next
example shows that this is not always the case and the relative
performance of the policies can be flipped.

\begin{example}[$\pi^g$ can be better than $\pi^G$]
  \label{ex:pi_g_v_pi_G}
  Consider a system at $t = 0$ with $J = 2$ and $N = 1$. For some
  fixed $\delta \in \left(0, \frac{1}{2}\right)$, let job 1 be
  characterized by
  $$\sigma_1 = 1,\quad v_1(t) = (1 - \delta)\Ind_{\set{t \leq 1}}.$$
  Let job 2 be characterized by
  $$\sigma_2 = 2, \quad v_2(t) = \Ind_{\set{t \leq 3}}.$$
  The policy $\pi^G$ will schedule job 2 and then job 1 so $V_0^G(s) =
  1$. In contrast, $\pi^g$ will schedule job 1 and then job 2 so
  $V_0^g(s) = 2 - \delta$. Therefore, $V_0^g(s) > V_0^G(s)$.
\end{example}

In contrast with Example~\ref{ex:pi_G_v_pi_g},
Example~\ref{ex:pi_g_v_pi_G} demonstrates the benefits of maximizing
the immediate reward rate rather than maximizing the immediate
reward. The distribution in Example~\ref{ex:pi_G_v_pi_g} is somewhat
pathological -- a service time like $\sigma_1$ in
Example~\ref{ex:pi_G_v_pi_g} is unlikely to arise in a most
applications. However, to give a rigorous performance guarantee that
holds for all service time distributions and value decay functions, we
need to consider such situations.

\begin{example}[$\pi^{EDF}$ can be better than $\pi^g$ and $\pi^G$ and vice-versa]
  \label{ex:EDF_v_g}
  Consider a system at $t = 0$ with $J = 2$ and $N = 1$. For $\eps \in
  (0, 1)$, define $\sigma$ as
  $$
  \sigma = \left\{
    \begin{array}{ll}
      1, & w.p.\quad \eps \\
      2, & w.p.\quad 1 - \eps \\
    \end{array}
  \right.
  $$
  and let $\sigma_1 \eqD \sigma_2 \eqD \sigma$. Let job 1 be
  characterized by $v_1(t) = \Ind_{\set{t \leq 1}}$ and job 2 be
  characterized by $v_2(t) = \Ind_{\set{t \leq 2}}$. The EDF policy
  will schedule job 1 and then job 2. Both jobs complete with
  probability $\eps^2$ and with probability $\eps(1 - \eps)$ we only
  complete job 1. Hence, $V_0^{EDF}(s) = \eps^2 + \eps$. In contrast,
  $\pi^g$ and $\pi^G$ will schedule job 2 and then job 1. Hence,
  $V_0^g(s) = V_0^G(s) = 1$.

  We can use the quadratic formula to show that if $\eps < (\sqrt{5} -
  1)/2$ then $V_0^{EDF}(s) < V_0^g(s) = V_0^G(s)$, if $\eps =
  (\sqrt{5} - 1)/2$ then $V_0^{EDF}(s) = V_0^g(s) = V_0^G(s)$, and if
  $\eps > (\sqrt{5} - 1)/2$ then $V_0^{EDF}(s) > V_0^g(s) = V_0^G(s)$.
\end{example}

\begin{remark}
  Note that in Example~\ref{ex:EDF_v_g}, $\pi^{EDF}$ only outperforms
  the other heuristics when the service times are nearly constant --
  in this example, when $\P(\sigma = 1) = \eps$ is qualitatively
  large. Otherwise, $\pi^{EDF}$ can perform arbitrarily poorly.  In
  contrast, $\pi^g$ and $\pi^G$ are insensitive to the value of
  $\eps$.
  \label{rem:EDF_v_g}
\end{remark}

These examples have demonstrated that each heuristic may be valuable
in different situations. To distinguish the heuristics, we now present
performance guarantees that hold for all service time distributions
(with finite mean) and all reward decay functions. Our performance
guarantees will be of the following form:

\begin{definition}
  Given a policy $\pi \in \Pi$, we say that $\pi$ is an
  $\alpha$-approximation if
  $$
  V_t^*(s) \leq \alpha V_t^\pi(s)
  $$
  for all $s$ and $t$. An optimal policy $\pi^*$ is a
  $1$-approximation and if a sub-optimal $\pi$ is an
  $\alpha$-approximation then $\alpha > 1$.
\end{definition}

We can think of $V_t^*(s)$ as the maximum amount of reward that is
available. If we used an optimal policy, we would be able to attain
all of this reward.  More generally, if a policy $\pi$ is an
$\alpha$-approximation, then we have a guarantee that $\pi$ will
attain a fraction $1/\alpha$ of the possible reward. Note that if
$\pi$ is an $\alpha$-approximation then it is also an
$\alpha'$-approximation for any $\alpha' \geq \alpha$.

\begin{theorem}
  \label{thrm:pi_G}
  Define
  $$
  \sigma_{max} = \max_{j \in \Jcal}\sigma_j \quad\text{and}\quad
  \sigma_{min} = \min_{j \in \Jcal}\sigma_j .$$
  Then $\pi^G$ is a $(1 +
  2\E[\sigma_{max}/\sigma_{min}])$-approximation.
\end{theorem}

The term $\E[\sigma_{max}/\sigma_{min}]$ shows that $\pi^G$ is
sensitive to the heterogeneity of the service times. If the service
times are deterministically equal, then $\sigma_{max}/\sigma_{min} =
1$, but typically, $\sigma_{max}/\sigma_{min} > 1$. When the service
times are deterministically equal, Theorem~\ref{thrm:pi_G} tells us
that $\pi^G$ is a $3$-approximation. As the gap between the largest
service time and the smallest service time widens, this guarantee
becomes weaker.

\begin{theorem}
  \label{thrm:pi_g}
  Define
  $$
  \Delta%
  = \frac{\E[\max_{j \in \Jcal} \sigma_j]}{\min_{j \in
      \Jcal}\E[\sigma_j]}%
  = \frac{\E[\sigma_{max}]}{\min_{j \in \Jcal}\E[\sigma_j]}%
  .$$ Then $\pi^g$ is a $(2 + \Delta)$-approximation.
\end{theorem}

The $\Delta$ term shows that $\pi^g$ is also sensitive to the
heterogeneity of the service times but in a different way. Note that
the performance guarantee for $\pi^g$ involves the minimum of the
expected services times ($\min_j \E[\sigma_j]$) while the performance
guarantee for $\pi^G$ involves the pointwise minimum ($\sigma_{min} =
\min_j \sigma_j$). The pointwise minimum is more sensitive to the
underlying service time distributions and this makes the performance
guarantee for $\pi^g$ somewhat more robust than the peformance
guarantee for $\pi^G$. However, we see a similar trend as with
$\pi^G$: when the service times are deterministically equal, $\pi^g$
is a $3$-approximation and as the service times become more
heterogeneous this performance guarantee weakens.

\begin{proposition}
  \label{prop:2approx}
  When $\E[\sigma_1] = \E[\sigma_2] = \cdots = \E[\sigma_J]$ then
  $\pi^g$ and $\pi^G$ are the same policy. Furthermore, when the
  service times are identically distributed, both $\pi^g$ and $\pi^G$
  are $2$-approximations.
\end{proposition}

This proposition shows that the performance guarantees in
Theorem~\ref{thrm:pi_G} and Theorem~\ref{thrm:pi_g} are not tight. In
the ``best'' case when the service times are deterministically equal,
the previous theorems told us that $\pi^g$ and $\pi^G$ were
$3$-approximations. However, the proposition tells us that they are
actually $2$-approximations. Intuitively, it is quite reasonable that
$\pi^g$ and $\pi^G$ have better guarantees when the services times are
independent and identically distributed (IID). Indeed, when there is
less heterogeneity amongst the jobs, scheduling decisions matter less
because the jobs are less distinguishable and hence, a greedy
heuristic should perform better. Our next example demonstrates that
when the services times are IID, the performance guarantee provided by
Proposition~\ref{prop:2approx} is tight. In other words, when the
service times are IID, $2$ is the smallest $\alpha$ such that $\pi^g$
and $\pi^G$ are $\alpha$-approximations.  That said, it is still unknown whether the bounds in Theorem~\ref{thrm:pi_G} and Theorem~\ref{thrm:pi_g} are tight under non-IID service time distributions. 

\begin{example}[Proposition~\ref{prop:2approx} is tight]
  Consider a system at $t = 0$ with $J = 2$ and $N = 1$. Let $\sigma_1
  = \sigma_2 = 1$ with probability 1. Fix $\eps \in (0, 1)$. The jobs
  are then distinguished by their value decay functions:
  $$v_1(t) = (1 - \eps)\Ind_{\set{t \leq 1}}, \quad v_2(t) = 1$$
  In this case, $\pi^g$ and $\pi^G$ will schedule job 2 and then job
  1. Hence, $V_0^g(s) = V_0^G(s) = 1$. On the other hand, an optimal
  policy will schedule job 1 and then job 2 which gives us $V_0^*(s) =
  2 - \eps$. Therefore, for any $\eps \in (0, 1)$ we have that
  $V_0^*(s) = (2 - \eps)V_0^g(s) = (2 - \eps)V_0^G(s)$. We can make
  $\eps$ arbitrarily small so the bound in Proposition~\ref{prop:2approx}
  is \emph{tight}.
\end{example}

Now we turn our attention to the EDF policy. Recall that $\pi^{EDF}$
is well-defined whenever the value decay functions reach zero in
finite time.  However, EDF policy does not make use of the service
time distributions, so that its performance may be arbitrarily bad for
heterogenous service time distributions. Indeed, even for the simpler
case of IID service times, Example~\ref{ex:EDF_v_g} showed that
$\pi^{EDF}$ can span the gamut from achieving nearly zero of the
possible reward to being optimal as the underlying service
distribution varies. In light of this, we focus our performance
analysis of EDF when the service distributions are IID (we relax this
assumption in our numeric experiments in Section \ref{sec:peva}). We
have the following performance guarantee:

\begin{theorem}
  \label{thrm:pi_EDF}
  Assume that the service times are identically distributed and define
  $$M = \max_{j,t}\set{\E[v_j(t + \sigma_j)]} %
  \quad\text{and}\quad %
  m = \min_{j,t}\set{\E[v_j(t + \sigma_j)] : \E[v_j(t + \sigma_j)] > 0}.$$
  Then $\pi^{EDF}$ is a $(1 + M/m)$-approximation.
\end{theorem}

Note that this performance guarantee is not very strong, in
general. However, when we consider reward functions which have hard
deadlines, it can be slightly refined:

\begin{corollary}
  \label{cor:pi_EDF}
  Assume that the service times are identically distributed and that
  $v_j(t) = \Ind_{\set{t \leq d_j}}$ for fixed $d_j$. Let $p_{min} =
  \min_{t}\set{F(t) : F(t) > 0}$. Then $\pi^{EDF}$ is a $(1 +
  1/p_{min})$-approximation.
\end{corollary}

Unfortunately,  the guarantees in Theorem~\ref{thrm:pi_EDF} and
Corollary~\ref{cor:pi_EDF} are quite weak. Generally, $m$ can be quite
small and hence $(1 + M/m)$ will be quite large. In the case of IID
service times, $\pi^g$ and $\pi^G$ were $2$-approximations regardless
of the underlying service time distribution. In contrast, we see that
the performance guarantee for $\pi^{EDF}$ is very sensitive to the
underlying distribution. If we focus on the situation in
Corollary~\ref{cor:pi_EDF}, note that when the service times are
deterministically equal to a constant, $p_{min} = 1$ and $\pi^{EDF}$
is a $2$-approximation. As the service time distributions become
stochastic, this guarantee is weakened. While this performance
guarantee may seem weak, it actually tight:

\begin{remark}[Theorem~\ref{thrm:pi_EDF} is tight]
  Example~\ref{ex:EDF_v_g} also demonstrates that the performance
  guarantee for $\pi^{EDF}$ is tight. We showed that $V_0^{EDF}(s) =
  \max\set{\frac{1}{\eps^2 + \eps}, 1}V_0^*(s)$. Note that in this
  case $1 + M/m = 1 + 1/\eps$. For $\eps < (\sqrt{5} - 1)/2$,
  $$
  \frac{V^*_0(s_0)/V^{EDF}_0(s_0)}{1 + M/m}
  = \frac{V^*_0(s_0)/V^{EDF}_0(s_0)}{1 + 1/\eps}
  = \frac{\frac{1}{\eps^2 + \eps}}{1 + \frac{1}{\eps}}
  = \frac{\eps}{\eps(\eps + 1)^2}
  = \frac{1}{(1 + \eps)^2}.
  $$
  Therefore, the ratio of the actual performance and the performance
  guarantee tends to 1 as $\eps \goesto 0$. In this sense, even though
  the performance can be arbitrarily bad, the performance guarantee is
  asymptotically tight.
\end{remark}

While this performance guarantee seems to suggest that the EDF policy
has weak performance, recall that this is not necessarily the case.
As shown by \cite{Argon_Triage_2008}, prioritizing time-critical jobs
can be optimal in special cases. In addition, Example~\ref{ex:EDF_v_g}
demonstrates that the EDF policy can outperform both the greedy and
rate greedy policies. We will see in our simulations that while the
EDF policy will not generally perform well, it can be a high
performing scheduling policy in special cases.

Given these theorems and examples, we summarize the given performance
guarantees in Table~\ref{tab:summary}. Note that in the case of IID
service times, all of the performance bounds are tight. In addition,
Proposition~\ref{prop:ordering} shows that we can order these
performance guarantees.

\begin{table}[h]
  \TABLE%
  {A summary of the performance guarantees for each of the heuristic policies.  \label{tab:summary}}%
  {\renewcommand{\arraystretch}{1.5}
    \begin{tabular}{|l||c|c|c|}
      \hline
      &  \multicolumn{3}{c|}{Heuristic}\\
      \cline{2-4}
      Service time distributions & $\pi^G$ & $\pi^g$ & $\pi^{EDF}$\\
      \hline
      Independent &
      $1 + 2\E[\sigma_{max}/\sigma_{min}]$&
      $2 + \Delta$&
      None available\\
      \hline
      Independent and Identically Distributed  &
      $2$ (tight)&
      $2$ (tight)&
      $1 + M/m$ (tight)\\
      \hline
    \end{tabular}
  }%
  {In each cell, we give a value for $\alpha$ such that the policy in question is an $\alpha$-approximation. Note that for IID service times, our performance guarantees are tight.}
\end{table}

\begin{proposition}
  \label{prop:ordering}
  The performance bounds can be ordered as follows:
  $$
  1 + 2\E\left[\frac{\sigma_{max}}{\sigma_{min}}\right] \geq 2 + \Delta;\quad %
  1 + \frac{M}{m} \geq 2$$
\end{proposition}

Proposition~\ref{prop:ordering} tells us that in the non-IID case,
$\pi^g$ has a better performance guarantee than $\pi^G$. In the IID
case, $\pi^g$ and $\pi^G$ each have better performance guarantees than
$\pi^{EDF}$. It appears that $\pi^g$ is better than $\pi^G$ and in the
IID case, both are better than $\pi^{EDF}$.  However, our examples
have shown that this is not always the case -- each of the policies
can attain a higher expected value than the other policies depending
on the situation. The examples were constructed to illustrate this
fact and so more numerical experiments are needed to evaluate the
performance of the different heuristics. 

Before conducting this numerical performance evaluation, we briefly
comment on the proofs of the performance guarantees (which can be
found in the appendices). The proofs are structurally similar and this
structure can potentially be leveraged for proving performance
guarantees for other myopic heuristics. The appendices illustrate how
after proving one performance guarantee we can modify certain bounds
to prove each subsequent performance guarantee. This opens the door
for exploring a multitude of other myopic scheduling heuristics that
might be of interest in other applications. The details of this
overarching proof structure is described in more detail in the
appendices.

\section{Performance Evaluation}\label{sec:peva}
We have given performance guarantees for each heuristic and have shown
that some of these bounds are  tight. These performance bounds
hold for arbitrary systems but we have seen in some simple examples
that the relative performance of the different policies can depend
on the system parameters. The system is parameterized by $N$, $J$,
$\set{F_j(\cdot)}_{j \in \Jcal}$, and $\set{v_j(\cdot)}_{j \in \Jcal}$
so even small problems have a high dimensional parameter space. Though
motivated primarily by patient scheduling in mass casualty incidents,
this model is broad enough to encompass several application
areas. Consequently, we opt to take two complementary approaches to
numerically explore this space:
\begin{enumerate}
\item We first focus on a handful of representative distributions and
  value decay functions which could be of potential interest to a
  variety of applications. We consider some relatively small problems
  in which we can compute $\pi^*$ thus allowing us to compare $\pi^*$,
  $\pi^g$, $\pi^G$, and $\pi^{EDF}$ for different combinations of
  service distributions and value decay functions. We also consider
  larger problems in which computing $\pi^*$ is not computationally
  tractable. In these larger cases, we simply compare the performance
  of each heuristic to the others.
\item We then conduct simulations in which the value decay functions
  and service time distributions would be of interest in healthcare
  operations management. Our model does not consider all of the
  details of a hospital, but our simulations do show how our results could be
  applied to patient scheduling in mass casualty scenarios.
\end{enumerate}

\subsection{General Numerical Experiments}
\begin{figure}[h]
  \FIGURE%
  {\includegraphics*[width=1.0\textwidth]{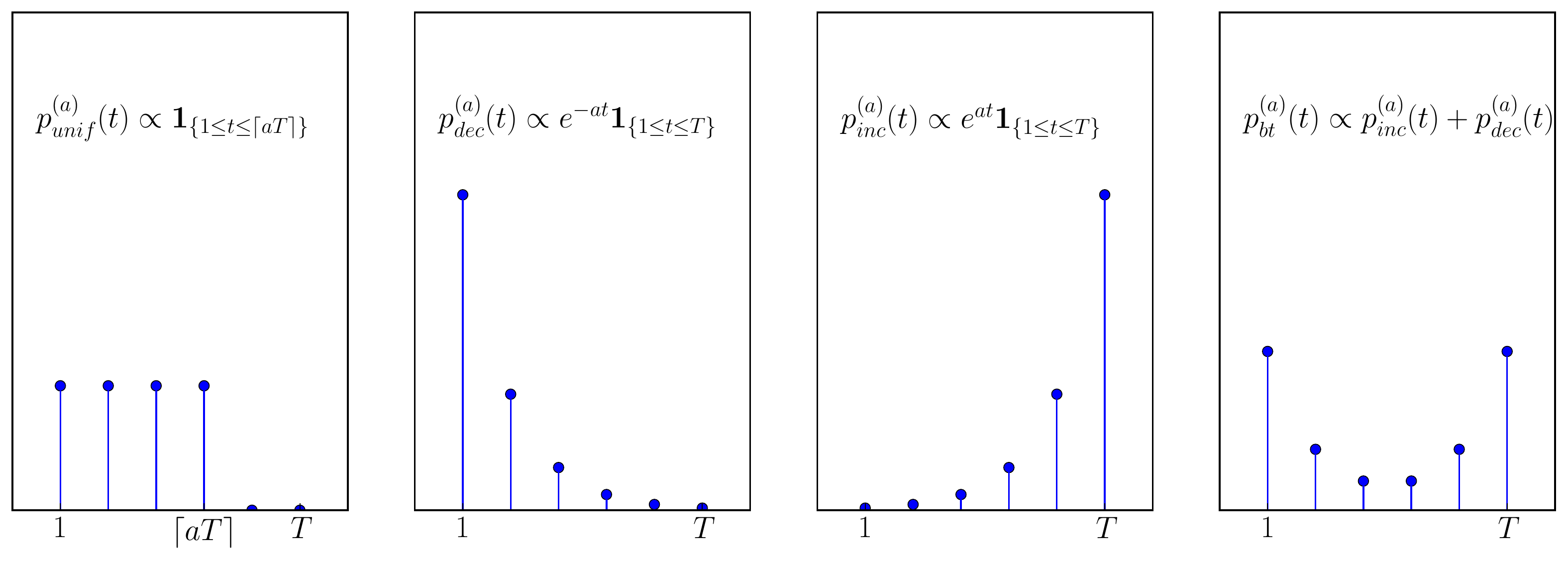}}%
  {Service Time Distributions. \label{fig:pmfs}}%
  {We consider four probability mass functions on $\set{1, \cdots, T}$
    corresponding to four different service time distributions each of
    which is parameterized by a value $a \in [0, 1]$. The first is
    uniform, the second is exponentially decreasing, and the third is
    exponentially increasing. The fourth probability mass function is
    a ``bathtub'' curve.}
\end{figure}

We consider four kinds of service time distributions, the probability
mass functions (PMFs) of which are depicted in
Figure~\ref{fig:pmfs}. We consider a uniform PMF, a exponentially
decreasing PMF, an exponentially increasing PMF, and a ``bathtub''
shaped PMF, each of which is parameterized by a value $a \in [0, 1]$
which can be used to adjust the mean and/or variance of the
distribution. The uniform PMF models the situation in which we do not
know anything about the job service time other than an upper bound
(i.e. $\lceil aT \rceil$) and a lower bound (i.e. $1$) and hence
choose the ``most random'' distribution (i.e. the maximum entropy
distribution). The increasing (decreasing) PMF models the situation in
which we believe the job is likely to complete service in a short
(long) period of time. Note that the decreasing PMF is a truncated
geometric PMF with parameter $(1 - e^{-a})$ and so for long time
horizons the decreasing PMF will be a good approximation for the
geometric PMF. Note that this is the discrete-time analog to the
exponential distribution which was used by \cite{Dalal_Impatient_2005}
when studying ``impatient'' users. The bathtub PMF is a bi-modal
distribution which models the situation in which we believe the job
will likely complete in either a short or long period of time but we
aren't sure which. This is conceptually similar to how bathtub curves
are used to model failure rates in reliability models
\citep{Xie_Bathtub_1996}.

\begin{figure}[h]
  \FIGURE%
  {\includegraphics*[width=1.0\textwidth]{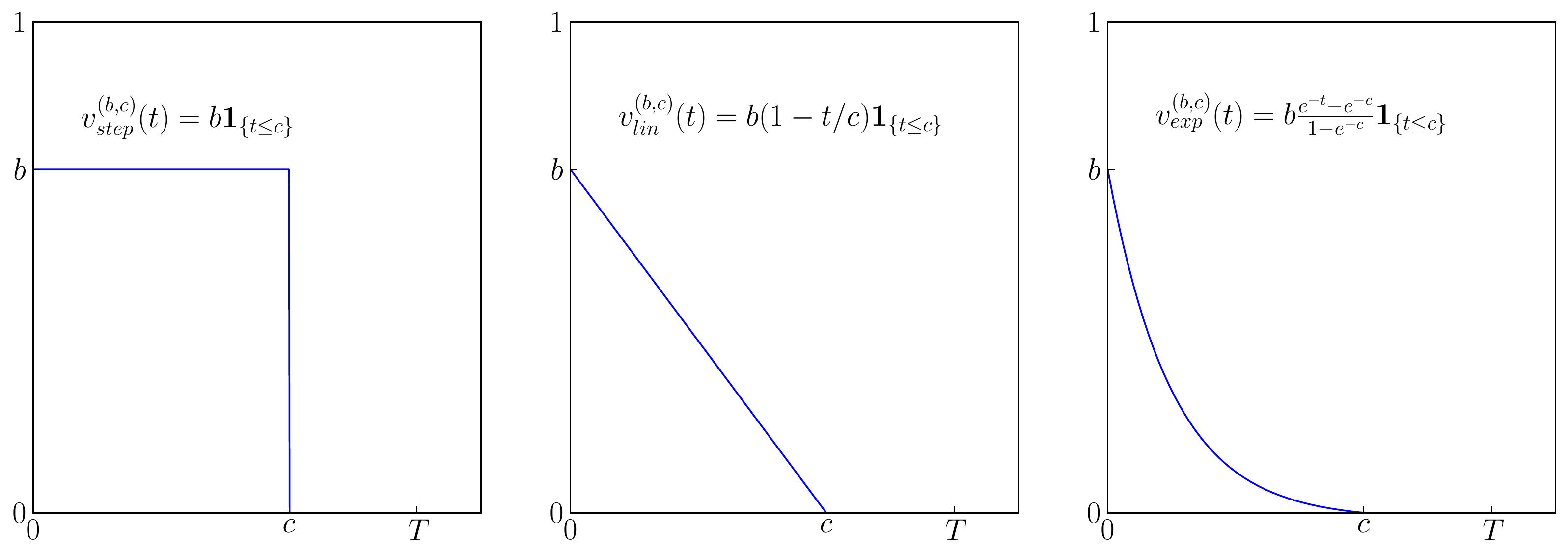}}%
  {Job Value Decay Functions. \label{fig:decays}}%
  {We consider three job value decay functions, each parameterized by
    a value $b \in [0, 1]$ and a deadline $c \in \set{1, \hdots,
      T}$. The first is a step function which can be used to model
    systems with job deadlines. The second is a linear decay function
    and the third is an exponential decay function.}
\end{figure}

As depicted in Figure~\ref{fig:decays}, we consider step-wise, linear,
and exponential value decay functions. Each of these functions is
parameterized by an initial value $b \in [0, 1]$ and a final deadline
$c \in \set{1, \hdots, T}$.  The step functions would be useful for
modeling jobs whose service requirements are characterized by
deadlines. If the internal value of the job decreases steadily, then a
linear decay function would be more appropriate. Finally, if there is
an incentive to complete service sooner and the job value will decay
rapidly, the exponential decay function would be the most appropriate
of these functions. Note that for large values of $c$, the exponential
value decay function is approximately the same as the function used by
\cite{Dalal_Impatient_2005} to model ``impatience'' amongst users.

\subsubsection{Small-Scale Problems: Comparisons to Optimal}~\\
We compare the policies in different situations by choosing all
combinations of the service time distributions and job value decay
functions. We also consider ``heterogeneous'' cases in which the job
service time distribution and/or family of value decay function is
chosen randomly and uniformly from the possible choices. In all cases,
the parameters $b$ and $c$ for each job value decay function are
chosen randomly and uniformly.  For a given combination of service
time distributions and value decay functions, let $\alpha_\pi =
V^*_0(s_0) / V_0^{\pi}(s_0)$ be the average of the ratio of the
optimal value and the value under policy $\pi$. For each myopic policy
and each combination, we estimate $\alpha_\pi$ and provide a standard
error by performing a Monte Carlo procedure with 1000 samples. We call
this estimate $\hat{\alpha}_\pi$. We give the results of the
comparison for a system with $J = 5$, $N = 2$, and a finite time
horizon of $T = 5$. Even though this may appear to be a small problem,
$\abs{\Scal} \approx 20,000$ and computing $\pi^*$ is non-trivial.

In Table~\ref{tab:J5_N2_T5}, we show the results for the case in which
we always take $a = 1$. Note that under this restriction, the service
times are IID, except in the case of ``Heterogeneous'', so that
$\pi^g$ and $\pi^G$ are the same policy.  For the first 3 rows,
$\pi^g$ and $\pi^G$ are both $2$-approximations (by
Proposition~\ref{prop:2approx}) while $\pi^{EDF}$ is a $(1 +
M/m)$-approximation (by Theorem~\ref{thrm:pi_EDF}).  In the
``Heterogeneous'' row, $\pi^g$ is a $(2 + \Delta)$-approximation (by
Theorem~\ref{thrm:pi_g}) and $\pi^G$ is a $(1 +
2\E[\sigma_{max}/\sigma_{min}])$-approximation (by
Theorem~\ref{thrm:pi_G}). In this case, $\pi^{EDF}$ does not have any
performance guarantee.

Table~\ref{tab:J5_N2_T5} shows that in many scenarios, $\pi^g$ and
$\pi^G$ both exhibit high performance. In fact, when the PMFs are
increasing, both $\pi^g$ and $\pi^G$ perform as well as the optimal
policy. When the service time distributions are heterogeneous, we
notice that $\pi^G$ consistently performs slightly better than
$\pi^g$. The difference is most substantial when the type of value
decay functions is also heterogeneous. This reinforces the intuition
we gleaned from Example~\ref{ex:pi_G_v_pi_g}: maximizing the expected
reward rate rather than the expected reward can lead to
suboptimal performance. We will further explore the performance differences
between $\pi^g$ and $\pi^G$ in  other numerical experiments.

In contrast to $\pi^g$ and $\pi^G$, $\pi^{EDF}$ does not perform very
well. In Table~\ref{tab:J5_N2_T5}, we see that $\hat{\alpha}_{EDF}$ is
very large, sometimes on the order of $10^5$. However, we do notice
some interesting trends. In particular, we see that for each row of
the table, $\pi^{EDF}$ performs best when the value decay is
step-wise. The EDF heuristic is motivated by deadlines so this matches
our intuition. When we further examine the column corresponding to
step-wise value decay, we see that $\pi^{EDF}$ performs best when the
PMFs are decreasing. This bolsters the intuition from
Example~\ref{ex:EDF_v_g} and Remark~\ref{rem:EDF_v_g}: $\pi^{EDF}$ can
perform well when there is high probability of completing each job in
a short amount of time. This is another phenomenon that we will
explore more in our other numerical experiments.

\begin{table}[h]
  \TABLE%
  {Performance of $\pi^G$, $\pi^g$, and $\pi^{EDF}$ compared to $\pi^*$ when $a = 1$ and $(b,c)$ is random.\label{tab:J5_N2_T5}}
  {\begin{minipage}{\textwidth}
      \setcounter{subtable}{0}
      \begin{subtable}{\textwidth}
        \centering
        \vspace{2mm}
        \begin{tabular}{|c|c|c|c|c|}
          \hline
          {}  & Step & Linear & Exponential & Heterogeneous\\
          \hline
          Uniform
          & 1.00615 (0.00058)
          & 1.00143 (0.00024)
          & 1.00073 (0.00013)
          & 1.00171 (0.00026)\\
          \hline
          Decreasing
          & 1.09513 (0.00303)
          & 1.01500 (0.00112)
          & 1.00445 (0.00051)
          & 1.05256 (0.00237)\\
          \hline
          Increasing
          & 1.00000 (0.00000)
          & 1.00000 (0.00000)
          & 1.00000 (0.00000)
          & 1.00000 (0.00000)\\
          \hline
          Bathtub
          & 1.01831 (0.00112)
          & 1.00478 (0.00055)
          & 1.00133 (0.00024)
          & 1.00997 (0.00090)\\
          \hline
          Heterogeneous
          & 1.03002 (0.00175)
          & 1.00465 (0.00063)
          & 1.00123 (0.00026)
          & 1.01682 (0.00140)\\
          \hline
        \end{tabular}
        \caption{$\hat{\alpha}_G$ $(\text{SE}(\hat{\alpha}_G))$}
      \end{subtable}

      \begin{subtable}{\textwidth}
        \centering
        \begin{tabular}{|c|c|c|c|c|}
          \hline
          {}  & Step & Linear & Exponential & Heterogeneous\\
          \hline
          Uniform
          & 1.00615 (0.00058)
          & 1.00143 (0.00024)
          & 1.00073 (0.00013)
          & 1.00171 (0.00026)\\
          \hline
          Decreasing
          & 1.09513 (0.00303)
          & 1.01500 (0.00112)
          & 1.00445 (0.00051)
          & 1.05256 (0.00237)\\
          \hline
          Increasing
          & 1.00000 (0.00000)
          & 1.00000 (0.00000)
          & 1.00000 (0.00000)
          & 1.00000 (0.00000)\\
          \hline
          Bathtub
          & 1.01831 (0.00112)
          & 1.00478 (0.00055)
          & 1.00133 (0.00024)
          & 1.00997 (0.00090)\\
          \hline
          Heterogeneous
          & 1.03087 (0.00171)
          & 1.00701 (0.00099)
          & 1.00673 (0.00101)
          & 1.03087 (0.00171)\\
          \hline
        \end{tabular}
        \caption{$\hat{\alpha}_g$ $(\text{SE}(\hat{\alpha}_g))$}
      \end{subtable}

      \begin{subtable}{\textwidth}
        \centering
        \vspace{2mm}
        \begin{tabular}{|c|c|c|c|c|}
          \hline
          {}  & Step & Linear & Exponential & Heterogeneous\\
          \hline
          Uniform
          & 2.49332 (0.04113)
          & 48.7847 (5.60725)
          & 289.450 (131.128)
          & 25.9185 (2.88039)\\
          \hline
          Decreasing
          & 1.14566 (0.00607)
          & 6.23741 (0.67474)
          & 9.43225 (0.67082)
          & 4.56169 (0.87904)\\
          \hline
          Increasing
          & 15.0676 (0.62802)
          & 200851. (48723.5)
          & 344279. (59059.1)
          & 69799.3 (27137.3)\\
          \hline
          Bathtub
          & 1.64460 (0.01817)
          & 15.3726 (1.21637)
          & 43.8726 (5.24929)
          & 11.3964 (1.36760)\\
          \hline
          Heterogeneous
          & 4.02893 (0.36334)
          & 2012.63 (997.432)
          & 3515.13 (2447.49)
          & 204.043 (81.5023)\\
          \hline
        \end{tabular}
        \caption{$\hat{\alpha}_{EDF}$ $(\text{SE}(\hat{\alpha}_{EDF}))$}
      \end{subtable}
    \end{minipage}
  }
  {
    We consider a system with $J = 5$ jobs, $N = 2$ processors,
    and a finite time-horizon of $T = 5$. We take $s_0$ to be the
    initial state in which all processors are free, no jobs have begun
    processing, and $t = 0$.  The columns in the table indicate the
    type of job value decay functions and the rows in the table
    indicate the kind of service time distribution. The parameters
    $(b, c)$ defining the job value decay functions are randomly
    chosen uniformly on $[0, 1] \times \set{1, \hdots, T}$ while we
    fix $a = 1$ for all service time distributions. When the column
    (row) is labeled ``heterogeneous'', the kind of value decay
    (service distribution) is chosen randomly and uniformly from the
    available kinds. Each scenario is repeated 1000 times and we
    report the average of $\alpha_\pi$ along with a standard error (to
    6 significant figures).
  }
\end{table}

In Table~\ref{tab:J5_N2_T5_mix}, we consider the case of when $a$,
$b$, and $c$ are all randomly chosen uniformly from their possible
values. In this case, $\pi^g$ is a $(2 + \Delta)$-approximation,
$\pi^G$ is a $(1 + 2\E[\sigma_{max}/\sigma_{min}])$-approximation, and
there is no known performance guarantee for $\pi^{EDF}$. Because all
three parameters are chosen randomly for each job, the jobs are more
diverse than they were in the previous numerical experiment. This
should ``stress'' the heurstics more because the optimal scheduling
choices are now less obvious.  Indeed, if we compare
Table~\ref{tab:J5_N2_T5} and Table~\ref{tab:J5_N2_T5_mix}, we
typically see larger values (i.e. lower performance) in
Table~\ref{tab:J5_N2_T5_mix}.  For example, in the ``Increasing'' row
of Table~\ref{tab:J5_N2_T5}, $\hat{\alpha}_g$ and $\hat{\alpha}_G$
were both equal to one but this is not the case for
Table~\ref{tab:J5_N2_T5_mix}. In both tables we notice the same
general trends that $\pi^g$ and $\pi^G$ perform quite well while
$\pi^{EDF}$ does not. Now that $\pi^g$ and $\pi^G$ are different for
all scenarios, we continue to see that $\pi^G$ performs slightly
better than $\pi^g$. The EDF policy performs quite poorly, but
performs best when the value decay is step-wise and the PMFs are
decreasing.

\begin{table}[h]
  \TABLE%
  {Performance of $\pi^G$, $\pi^g$, and $\pi^{EDF}$ compared to $\pi^*$ when $(a, b,c)$ is random.\label{tab:J5_N2_T5_mix}}
  {\begin{minipage}{\textwidth}
      \setcounter{subtable}{0}
      \begin{subtable}{\textwidth}
        \centering
        \vspace{2mm}
        \begin{tabular}{|c|c|c|c|c|}
          \hline
          {}  & Step & Linear & Exponential & Heterogeneous\\
          \hline
          Uniform
          & 1.04897 (0.002167)
          & 1.00768 (0.000844)
          & 1.00138 (0.000236)
          & 1.01387 (0.001245)\\
          \hline
          Decreasing
          & 1.04691 (0.002012)
          & 1.00788 (0.000758)
          & 1.00180 (0.000268)
          & 1.02233 (0.001544)\\
          \hline
          Increasing
          & 1.00127 (0.000230)
          & 1.00023 (0.000066)
          & 1.00009 (0.000031)
          & 1.00037 (0.000104)\\
          \hline
          Bathtub
          & 1.00041 (0.000089)
          & 1.00181 (0.000293)
          & 1.00011 (0.000035)
          & 1.00017 (0.000046)\\
          \hline
          Heterogeneous
          & 1.02546 (0.001638)
          & 1.00276 (0.000504)
          & 1.00076 (0.000199)
          & 1.00783 (0.000948)\\
          \hline
        \end{tabular}
        \caption{$\hat{\alpha}_G$ $(\text{SE}(\hat{\alpha}_G))$}
      \end{subtable}

      \begin{subtable}{\textwidth}
        \centering
        \begin{tabular}{|c|c|c|c|c|}
          \hline
          {}  & Step & Linear & Exponential & Heterogeneous\\
          \hline

          Uniform
          & 1.05640 (0.002226)
          & 1.00906 (0.000852)
          & 1.00760 (0.000962)
          & 1.01485 (0.001315)\\
          \hline
          Decreasing
          & 1.04330 (0.001952)
          & 1.00923 (0.001002)
          & 1.00518 (0.000558)
          & 1.02275 (0.001470)\\
          \hline
          Increasing
          & 1.00154 (0.000243)
          & 1.00078 (0.000174)
          & 1.00065 (0.000121)
          & 1.00128 (0.000217)\\
          \hline
          Bathtub
          & 1.00093 (0.000148)
          & 1.00181 (0.000293)
          & 1.00042 (0.000102)
          & 1.00047 (0.000110)\\
          \hline
          Heterogeneous
          & 1.02546 (0.001638)
          & 1.00461 (0.000682)
          & 1.00455 (0.000631)
          & 1.01150 (0.001229)\\
          \hline
        \end{tabular}
        \caption{$\hat{\alpha}_g$ $(\text{SE}(\hat{\alpha}_g))$}
      \end{subtable}

      \begin{subtable}{\textwidth}
        \centering
        \vspace{2mm}
        \begin{tabular}{|c|c|c|c|c|}
          \hline
          {}  & Step & Linear & Exponential & Heterogeneous\\
          \hline
          Uniform
          & 1.78906 (0.033164)
          & 33.3784 (6.97825)
          & 85.3505 (32.4023)
          & 95.5145 (75.0312)\\
          \hline
          Decreasing
          & 1.51489 (0.018046)
          & 14.4466 (1.63058)
          & 51.9981 (21.8992)
          & 6.62267 (0.605458)\\
          \hline
          Increasing
          & 7.08516 (0.334549)
          & 2467.51 (354.735)
          & 8653.56 (1422.52)
          & 9076.97 (6364.62)\\
          \hline
          Bathtub
          & 6.06017 (0.197840)
          & 44.1500 (7.09408)
          & 5034.91 (1911.05)
          & 831.657 (224.303)\\
          \hline
          Heterogeneous
          & 3.90997 (0.185022)
          & 552.404 (146.859)
          & 2386.36 (683.175)
          & 234.359 (93.5448)\\
          \hline
        \end{tabular}
        \caption{$\hat{\alpha}_{EDF}$ $(\text{SE}(\hat{\alpha}_{EDF}))$}
      \end{subtable}
    \end{minipage}
  }
  {
    We consider a system with $J = 5$ jobs, $N = 2$ processors, and a
    finite time-horizon of $T = 5$. We take $s_0$ to be the initial state
    in which all processors are free, no jobs have begun processing, and
    $t = 0$.  The columns in the table indicate the type of job value
    decay functions and the rows in the table indicate the kind of service
    time distribution. The parameters $(a, b, c)$ defining each job are
    randomly chosen uniformly on $[0,1] \times [0, 1] \times \set{1,
      \hdots, T}$. When the column (row) is labeled ``heterogeneous'', the
    kind of value decay (service distribution) is chosen randomly and
    uniformly from the available kinds. Each scenario is repeated 1000
    times and we report the average of $\alpha_\pi$ along with a standard
    error (to 6 significant figures).
  }
\end{table}

\subsubsection{Large-Scale Problems: Comparing the Heuristics to Each Other}
\begin{figure}
  \FIGURE
  {
    \begin{minipage}{\textwidth}
      \begin{subfigure}{0.48\textwidth}
        \includegraphics[width=\textwidth]{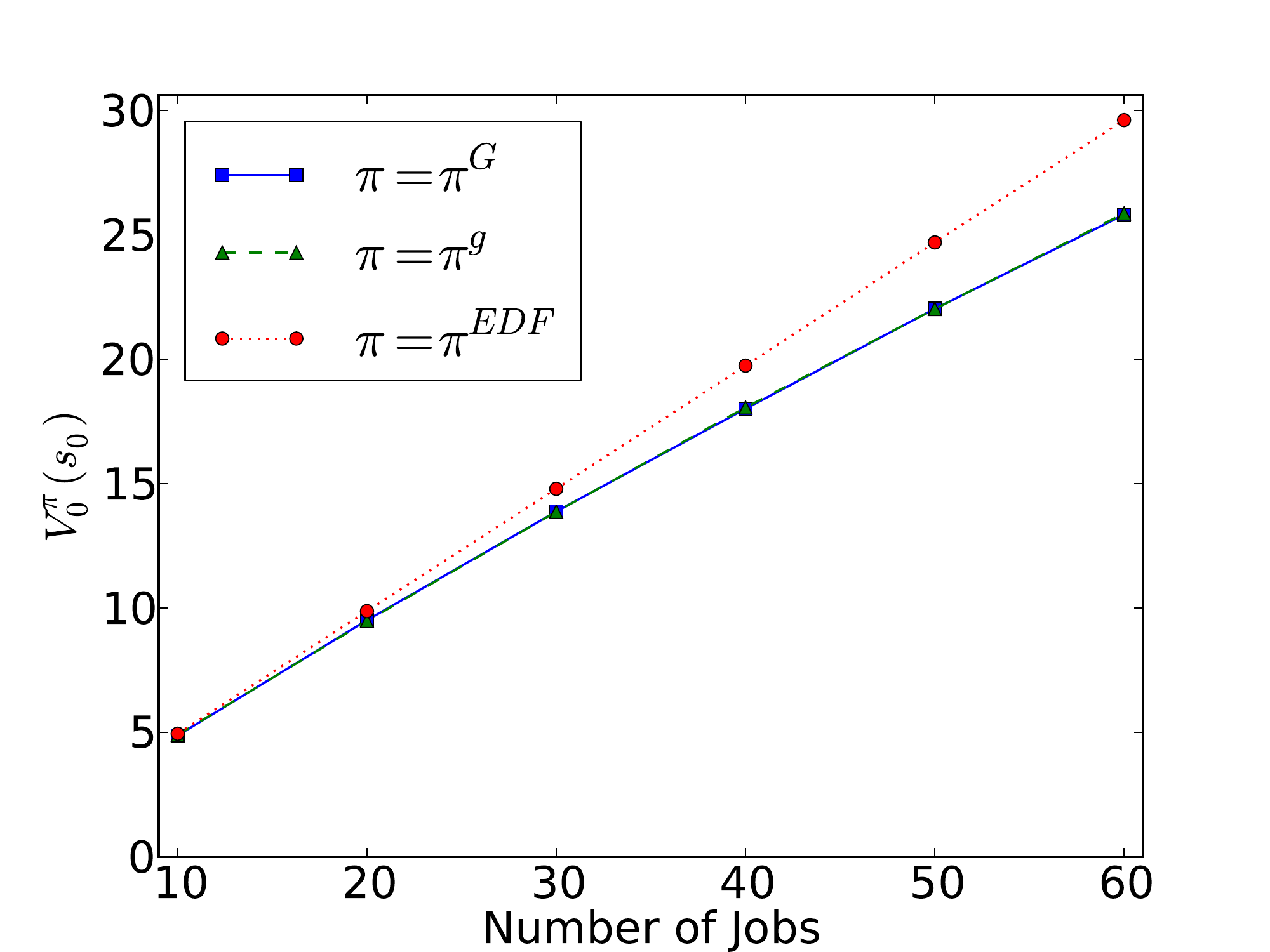}
        \caption{\footnotesize Step-wise decay, Geometric
          PMFs\label{fig:vary_jobs_sim_step_dec}}
      \end{subfigure}
      \begin{subfigure}{0.48\textwidth}
        \includegraphics[width=\textwidth]{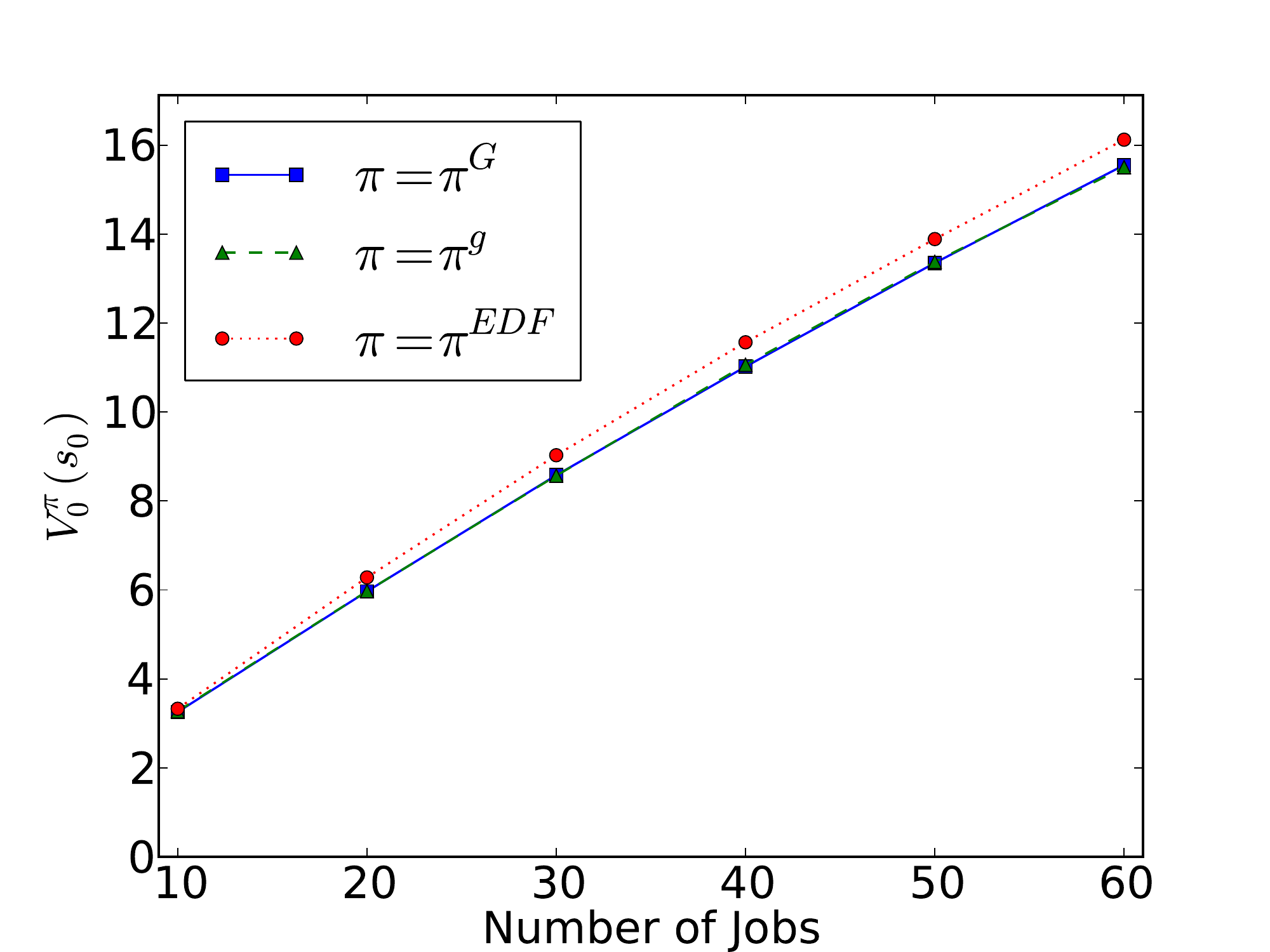}
        \caption{\footnotesize Heterogeneous decay, Geometric
          PMFs\label{fig:vary_jobs_sim_rand_dec}}
      \end{subfigure}\\
      \begin{subfigure}{0.48\textwidth}
        \includegraphics[width=\textwidth]{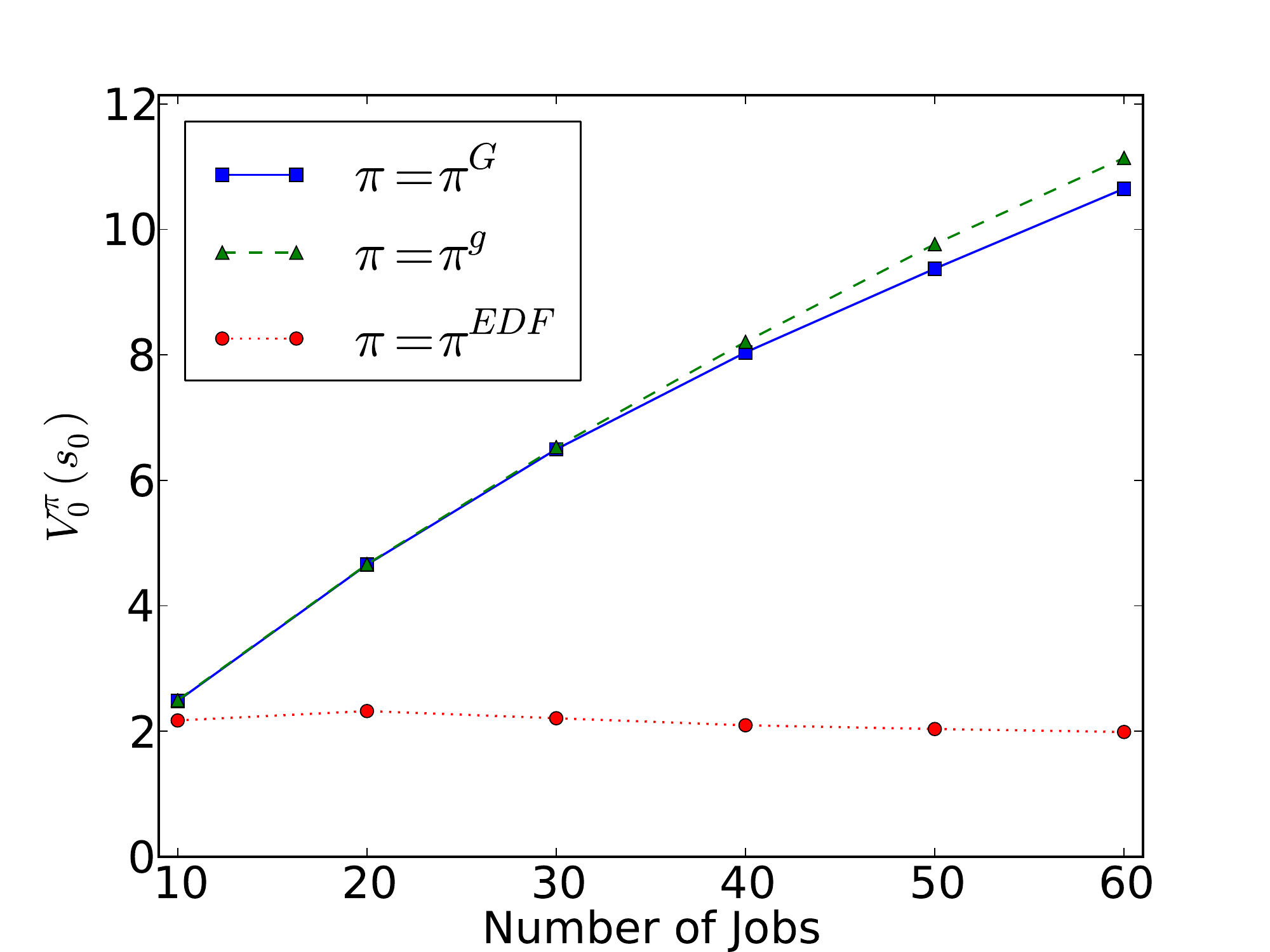}
        \caption{\footnotesize Step-wise  decay, Heterogeneous PMFs\label{fig:vary_jobs_sim_step_rand}}
      \end{subfigure}
      \begin{subfigure}{0.48\textwidth}
        \includegraphics[width=\textwidth]{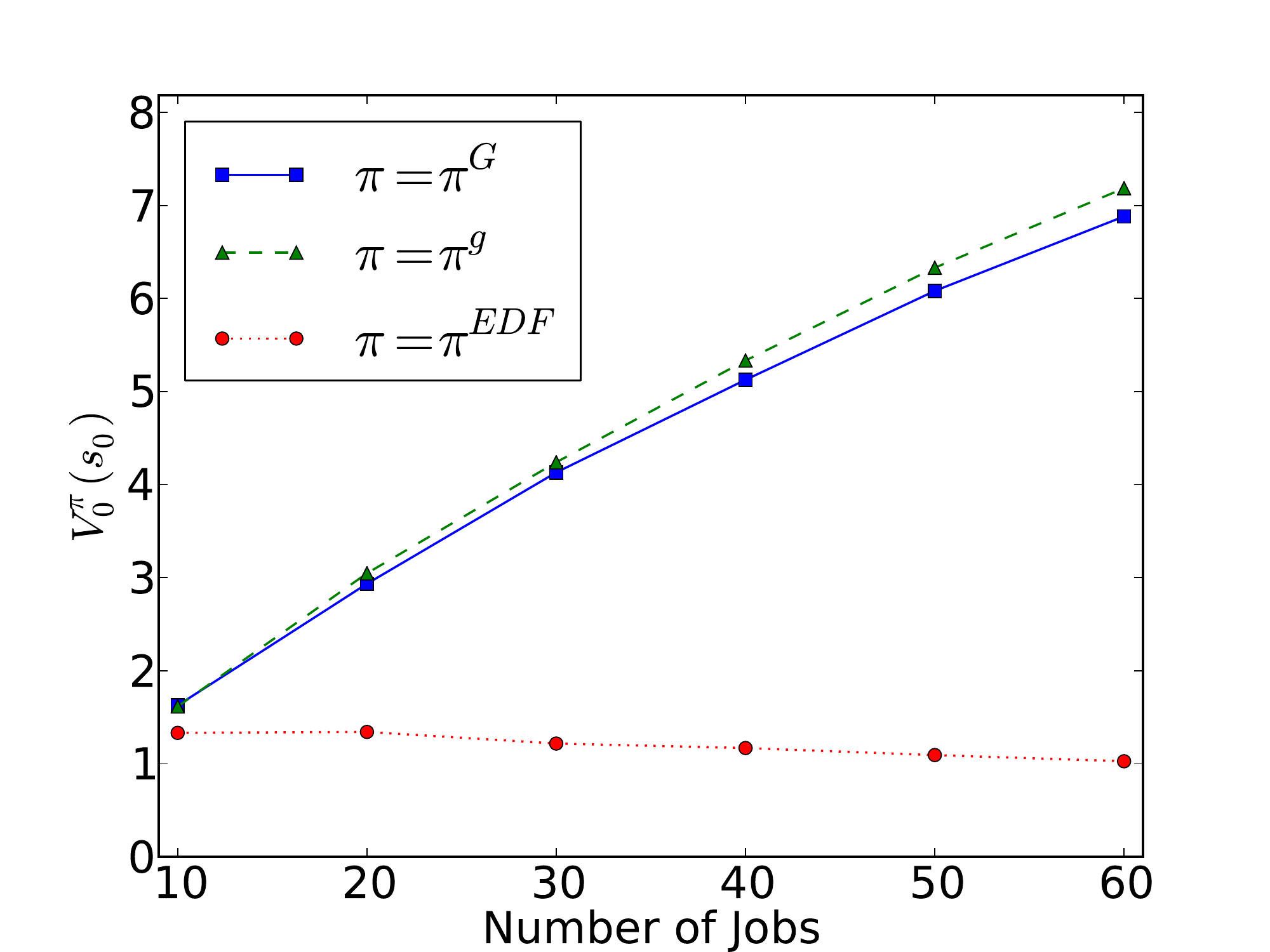}
        \caption{\footnotesize Heterogeneous  decay, Heterogeneous
          PMFs\label{fig:vary_jobs_sim_rand_rand}}
      \end{subfigure}
    \end{minipage}
  }%
  {Performance of $\pi^G$, $\pi^g$, and $\pi^{EDF}$ when $a = 1$ and
$(b,c)$ is random.\label{fig:vary_jobs_sim}}%
  {We compare the performance of the heuristics when $N = 5$
    and $T = 50$ while varying $J$. }
\end{figure}

\begin{figure}
  \FIGURE%
  {
    \begin{minipage}{\textwidth}
      \begin{subfigure}{0.48\textwidth}
        \includegraphics[width=\textwidth]{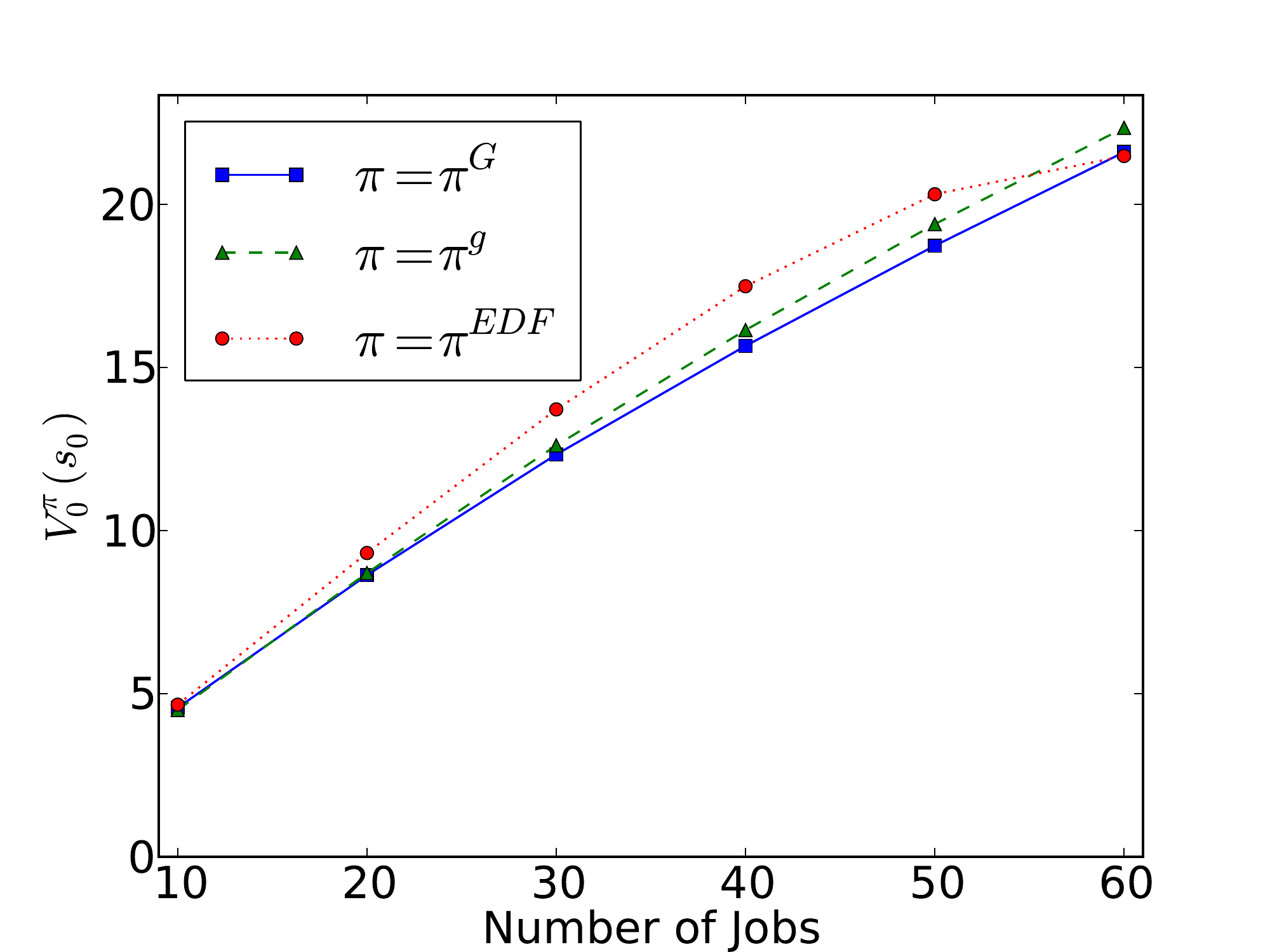}
        \caption{\footnotesize Step-wise decay, Geometric
          PMFs\label{fig:vary_jobs_sim_step_dec_mixed}}
      \end{subfigure}
      \begin{subfigure}{0.48\textwidth}
        \includegraphics[width=\textwidth]{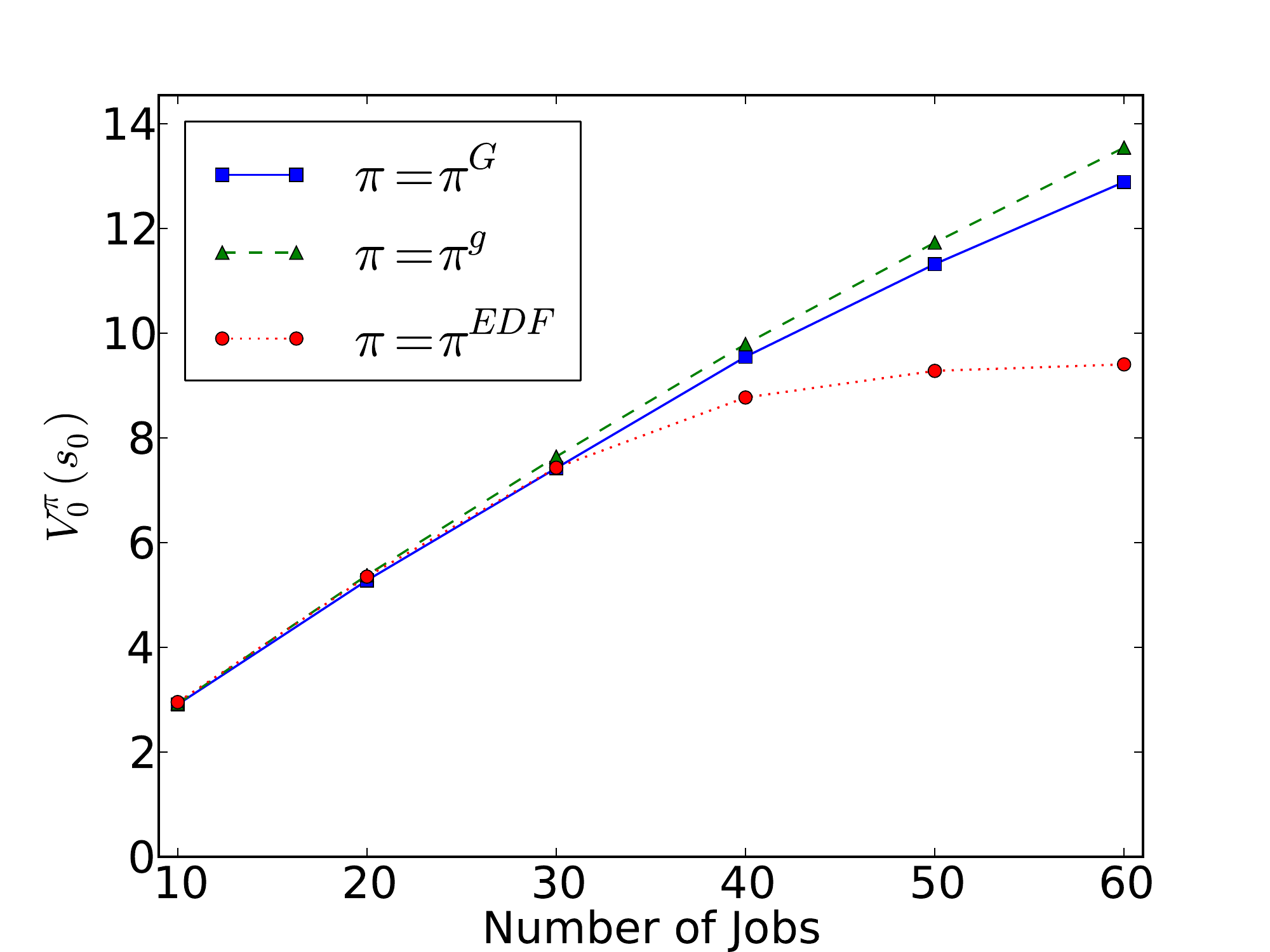}
        \caption{\footnotesize Heterogeneous decay, Geometric
          PMFs\label{fig:vary_jobs_sim_rand_dec_mixed}}
      \end{subfigure}\\
      \begin{subfigure}{0.48\textwidth}
        \includegraphics[width=\textwidth]{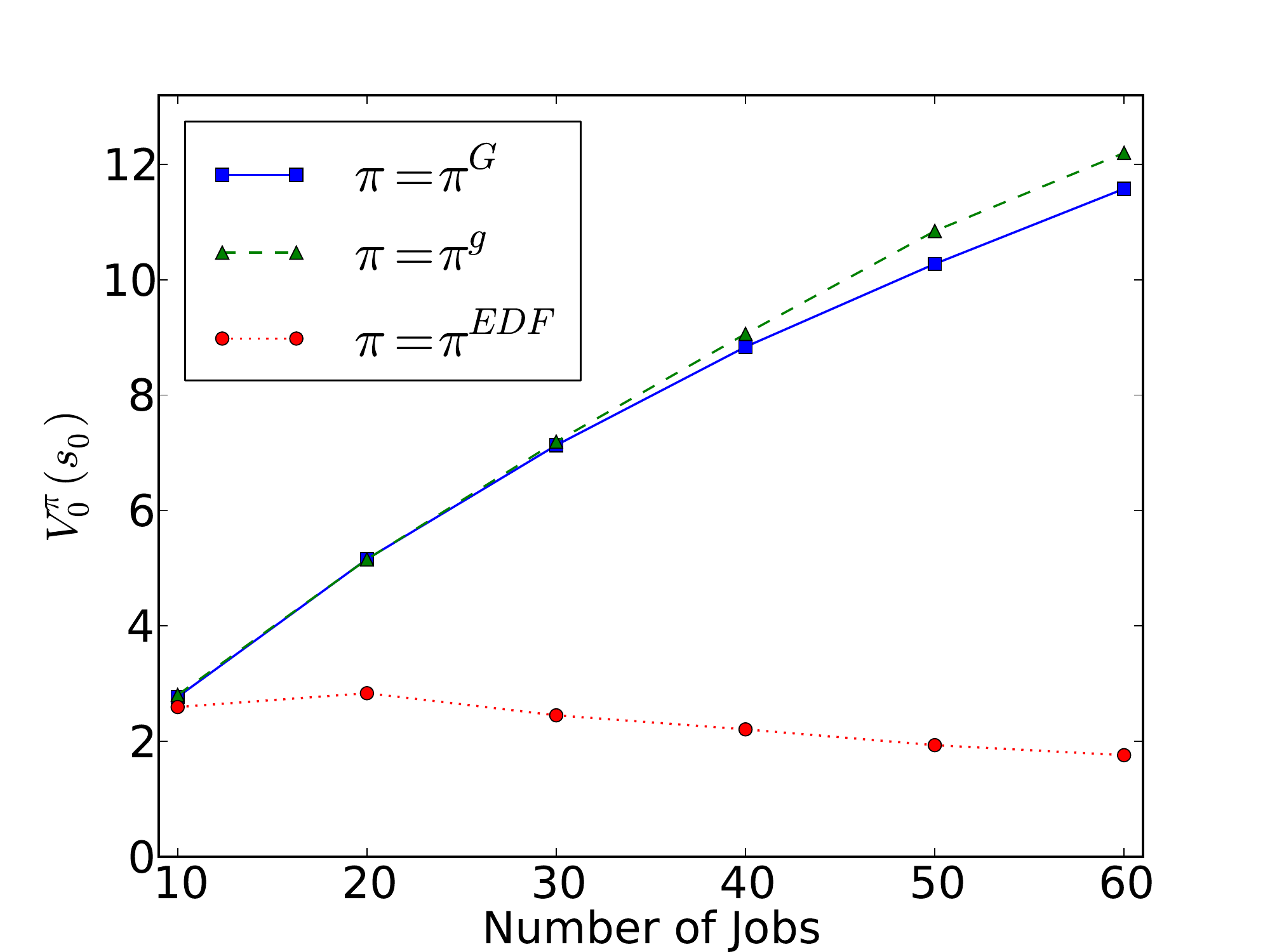}
        \caption{\footnotesize Step-wise decay, Heterogeneous
          PMFs\label{fig:vary_jobs_sim_step_rand_mixed}}
      \end{subfigure}
      \begin{subfigure}{0.48\textwidth}
        \includegraphics[width=\textwidth]{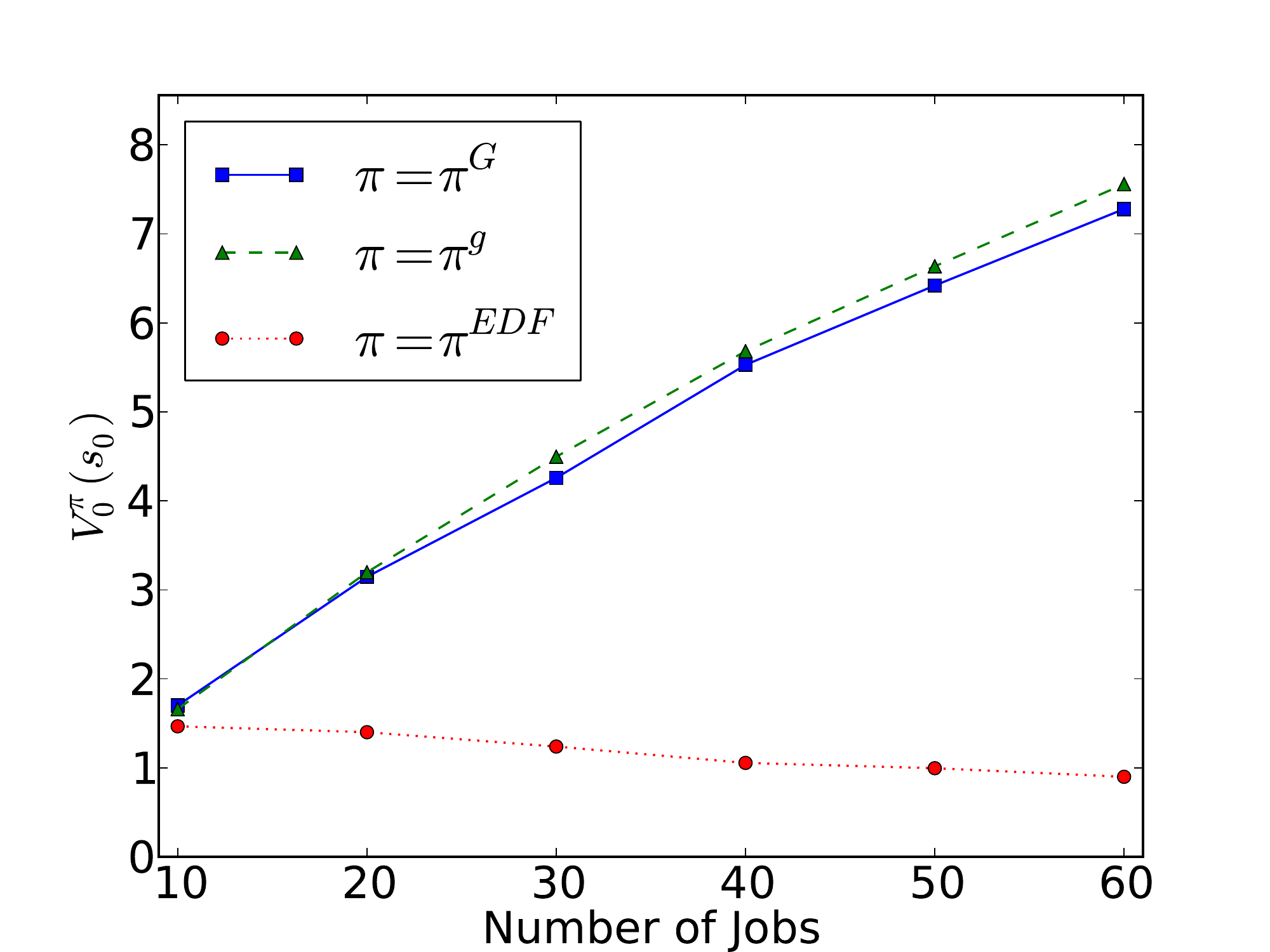}
        \caption{\footnotesize Heterogeneous decay, Heterogeneous
          PMFs\label{fig:vary_jobs_sim_rand_rand_mixed}}
      \end{subfigure}
    \end{minipage}
  }%
  {Performance of $\pi^G$, $\pi^g$, and $\pi^{EDF}$ when $(a, b,c)$ is
    random.\label{fig:vary_jobs_sim_mixed}}%
  {We compare the performance of the heuristics when $N = 5$ and $T =
    50$ while varying $J$. }
\end{figure}

~\\
We now consider larger problems in which computing $\pi^*$ is not
feasible. We consider a system with a finite time horizon of $T = 50$
and $N = 5$ while varying $J \in \set{10, 20, 30, 40, 50,
  60}$. Increasing $J$ corresponds to increasing the congestion of the
system. In a more congested system, the appropriate scheduling
decisions become both more critical and less obvious. The state space
is now so large that computing an optimal policy is not feasible and
exact performance evaluation of sub-optimal policies is also not
feasible. Instead, we randomly choose the system parameters and
simulate the system evolution under the various heuristic policies. We
report the average result of $1000$ simulations. We consider either
step-wise or heterogeneous value decay and either decreasing or
heterogeneous PMFs. Because $T = 50$, the decreasing PMF is ``almost''
a geometric PMF. Indeed, if $\sigma$ is a geometric random variable on
$\set{1, 2, \hdots}$ with parameter $(1 - 1/e)$, then $\P(\sigma > T)
\approx 10^{-16}$. As a result, in this section we will refer to the
decreasing PMF as a geometric PMF. Therefore, we are comparing the
special cases of deadlines and geometric service times against
heterogeneous value decay functions and heterogeneous service time
distributions. In every case, the parameters $b$ and $c$ which define
each job value decay function are chosen randomly. In
Figure~\ref{fig:vary_jobs_sim}, we consider when $a = 1$ and in
Figure~\ref{fig:vary_jobs_sim_mixed}, we consider when $a$ is also
chosen randomly.

In Figure~\ref{fig:vary_jobs_sim_step_dec} we have geometric PMFs and
step-wise value decay while in Figure~\ref{fig:vary_jobs_sim_rand_dec}
we have geometric PMFs and heterogeneous value decay. Note that in
these cases, $\pi^g$ and $\pi^G$ are equivalent. In both plots we see
that the value associated with each of the three heuristic policies
increases as the number of jobs increases. This shows that when the
service times are geometric, all three heuristics can manage
increasing congestion reasonably well. An interesting feature of these
plots is that $\pi^{EDF}$ outperforms $\pi^g$ and $\pi^G$. This
matches our intuition from Example~\ref{ex:EDF_v_g} and the
small-scale simulation from the previous section. In the previous
section, we saw that $\pi^{EDF}$ performed best when the PMFs were
(truncated) geometrics. Recall that in Example~\ref{ex:EDF_v_g}, when
$\P(\sigma = 1)$ is close to 1, $\pi^{EDF}$ outperforms $\pi^g$ and
$\pi^G$. Because $(1 - 1/e) \approx 0.63$, the probability of a job
completing in one time slot is indeed qualitatively close to 1. In
these examples, all of the jobs are very likely to finish in a short
amount of time and since $\pi^{EDF}$ prioritizes time-critical jobs,
$\pi^{EDF}$ does very well. Note that $\pi^g$ and $\pi^G$ also perform
well. Furthermore, when the value decay functions are heterogeneous,
the performance gap between $\pi^g/\pi^G$ and $\pi^{EDF}$ is quite
small. These plots show that while it may be slightly better to use
$\pi^{EDF}$ when the service times are short (with high probability),
$\pi^g$ and $\pi^G$ are both good options as well.

In Figure~\ref{fig:vary_jobs_sim_step_rand} we have heterogeneous PMFs
and step-wise value decay while in
Figure~\ref{fig:vary_jobs_sim_rand_rand} we have heterogeneous PMFs
and heterogeneous value decay. Note that in these cases, the
performance guarantee for $\pi^{EDF}$ does not apply because the
service times are not identically distributed. We see that $\pi^{EDF}$
performs quite poorly and the value associated with $\pi^{EDF}$ does
not increase as $J$ is increased. This demonstrates that with
heterogeneous PMFs, $\pi^{EDF}$ does not handle congestion well. With
heterogeneous PMFs, there are some jobs which will complete quickly,
but many of the jobs will not. The results in
Figure~\ref{fig:vary_jobs_sim} demonstrate that the EDF policy is
sensitive to the underlying service time distributions: $\pi^{EDF}$
will perform well when there is a high probability that the jobs will
each complete in a single time slot, but otherwise $\pi^{EDF}$ will
perform quite poorly. In contrast, both $\pi^g$ and $\pi^G$ perform
well in the heterogeneous environments and are able to gain a greater
reward as $J$ increases. In Figure~\ref{fig:vary_jobs_sim_step_rand},
we see that $\pi^g$ and $\pi^G$ perform nearly identically until $J >
30$. For $J > 30$, $\pi^g$ performs slightly better than $\pi^G$. In
Figure~\ref{fig:vary_jobs_sim_rand_rand}, this dichotomy becomes
evident when $J > 10$. This suggests that though $\pi^g$ and $\pi^G$
perform similarly for small problems, $\pi^g$ will be better when the
system is more congested. Furthermore, the benefit of $\pi^g$ over
$\pi^G$ is more evident when there is greater heterogeneity amongst
the jobs.

Figure~\ref{fig:vary_jobs_sim_mixed} (the case in which the parameter
$a$ is randomly chosen for each job rather than being fixed), shows
many of the same trends as Figure~\ref{fig:vary_jobs_sim} but gives us
additional insights into how heterogeneity in service times affects
each heuristic policy. First note that each subfigure in
Figure~\ref{fig:vary_jobs_sim_mixed} shows that $\pi^g$ outperforms
$\pi^G$ and that the difference in performance increases as $J$
increases. This demonstrates that for large numbers of jobs $\pi^g$ is
superior to $\pi^G$. In Figure~\ref{fig:vary_jobs_sim_step_dec} and
Figure~\ref{fig:vary_jobs_sim_rand_dec}, we saw that $\pi^{EDF}$
gained more rewards as $J$ increased. However, in
Figure~\ref{fig:vary_jobs_sim_step_dec_mixed} and
Figure~\ref{fig:vary_jobs_sim_rand_dec_mixed} we see that the value
associated with $\pi^{EDF}$ saturates as $J$ increases while the
values associated with $\pi^G$ and $\pi^g$ do not. This supports the
idea that $\pi^{EDF}$ is not nearly as robust as $\pi^g$ and
$\pi^G$. Although $\pi^{EDF}$ can perform well, it may be more prudent
to apply $\pi^g$ or $\pi^G$. Futhermore, for larger problems it may be
best to use $\pi^g$ rather than $\pi^G$.

\subsubsection{Rules-of-Thumb}~\\
Our numerical experiments and theoretical results have revealed
several insights into the general problem of non-preemptive scheduling
of jobs with decaying value. We summarize these insights with the
following rules-of-thumb:
\begin{itemize}
\item The rate greedy policy has the best performance guarantee.  The
  greedy policy has a better performance guarantee than EDF. In this
  sense, rate greedy policy is the most robust to changes in the
  underlying system parameters.

\item For problems with relatively short time horizons, the greedy
  policy performs best.

\item Despite the seemingly weak performance guarantee, EDF performs
  well under long time horizons when there is a high probability that
  the service times are short. In these situations, the rate greedy
  and greedy policies also perform very well, but EDF can perform even
  better.

\item When the service times are large with high probability
  and/or if the reward decay function cannot be characterized by deadlines,
  EDF performs poorly.  In these cases, it is better to use
  one of the greedy policies.

\item For problems with long time horizons and heterogenous service
  time distributions , the rate greedy policy performs slightly better
  than the greedy policy.  This difference is more pronounced when the
  reward decay functions are also heterogeneous.
\end{itemize}

\subsection{Patient Scheduling in a Disaster Scenario\label{ssec:patients}}
We now consider a patient scheduling problem where patients are
``jobs'' and operating rooms in a surgical center are ``servers''.  We
are concerned with the 24 hour period immediately after a mass
casualty incident as studies have shown that this is a critical period
for hospitals responding to mass casualty incidents
(e.g. \cite{Aylwin_London_2007}, \cite{Turegano_Madrid_2008,
Turegano_Madrid2_2008}). The random service times correspond to
uncertainty in procedure durations and the internal value decay
corresponds to the deterioration of patient health due to scheduling
delays. For example, recall the ``health score'' used by
\cite{Sacco_Triage_2005} to model the decline in patient health as
procedures are delayed. We consider a clearing system which is often
used to model mass casualty incidents (e.g. \cite{Argon_EORMS_2011}
and \cite{Chan_Burns_2013}). Our model does not explicitly consider
the details of a surgical operation (e.g. the pre-operative and
post-operative phases), but does address the key dilemma of scheduling
and prioritizing patients when there is a scarcity of operating
rooms. Sub-optimal operating room schedules can lead to delays even in
typical circumstances \citep{Wachtel_OR_2009} and a spike in demand
due to a disaster will only exacerbate this issue, so our model is
able to capture a primary operational concern.

For the service time distributions, we use lognormal
distributions. \cite{Strum_LogNormal_2000} showed that lognormal
distributions model surgical procedure times better than normal
distributions. Furthermore, \cite{Spangler_LogNormal_2004}
demonstrated how to fit the parameters of a lognormal distribution to
surgical data from hospitals. The use of lognormal distribution in
modeling surgical procedure times is now quite common
(e.g. \cite{Mihaylova_HealthcareMethods_2011}). We have elected to
calibrate our simulation according to surgical procedures because
these types of procedures are common when managing mass casualty
incidents, e.g. during civilian terrorist attacks
\citep{Frykberg_MCISurgery_2004} and in the aftermath of military
combat \citep{King_MASH_2005}.

Because we have a discrete-time model, we need to discretize the
lognormal density. Given parameters $\ell$, $m$, $s$, and a standard
normal random variable $Z$, $\sigma$ is lognormal if
\begin{equation}
  \sigma = \ell + e^{m + s Z}.
\end{equation}
We will assume that $\sigma$ is measured in minutes. Given a time
discretization $\delta$ and a number of time slots $T$, we can compute
a PMF $p_{ln}(\cdot)$ which approximates the density of $\sigma$. See
the appendix for details.

For each patient, we fix $\ell = 60$, randomly select $m$ from a
uniform distribution on $[1.0, 4.0]$, and randomly select $s$ from a
uniform distribution on $[1.0, 1.25]$.  With these parameters,
expected procedure times are on the order of two to three hours, and
the coefficient of variation for each procedure is between 0.1 and
1.5. This range for the coefficients of variation is motivated by
\cite{Spangler_LogNormal_2004} who showed that when fitting lognormal
random variables to surgical procedure times, nearly procedures
studied had coefficients of variation less than 1.5. While the study
by \cite{Spangler_LogNormal_2004} was not motivated by disaster
scenarios, we note that there is an inherent difficulty in fitting
statistical models to the types of procedures which arise during mass
casualty incidents. \cite{Frykberg_MCISurgery_2004} points out that
the procedures required during mass casualty incidents are
characterized by ``complex and difficult wounding patterns that are
not typically seen in routine practice'' and that the rare nature of
these procedures makes controlled statistical analysis difficult to
perform. Consequently, while the selected parameter ranges are
reasonable and somewhat plausible, one would need to make more
judicious parameter choices in order to model specific scenarios.
Incorporating expert knowledge can be useful when a data driven
approach is not feasible and as mentioned before, one could apply the
Delphi method \citep{Linstone_Delphi_1975} to build models for
specific disasters and injury types.

We consider a 24 hour period discretized into 10 minute time slots so
that $\delta = 10$ and $T = 144$. We assume there are 6 operating
rooms (the average number of operating rooms in a hospital in the
United States according to \cite{Gallup}) and that the medical
resources are sufficient to complete operations at a constant rate. We
assume that the time between procedures is negligible. If we assume
that preparations such as the application of anesthesia are included
in the service time distribution (as is done in
\cite{Spangler_LogNormal_2004}), then this is a fairly benign
assumption.

\begin{figure}[h]
  \FIGURE%
  {\includegraphics*[width=0.5\textwidth]{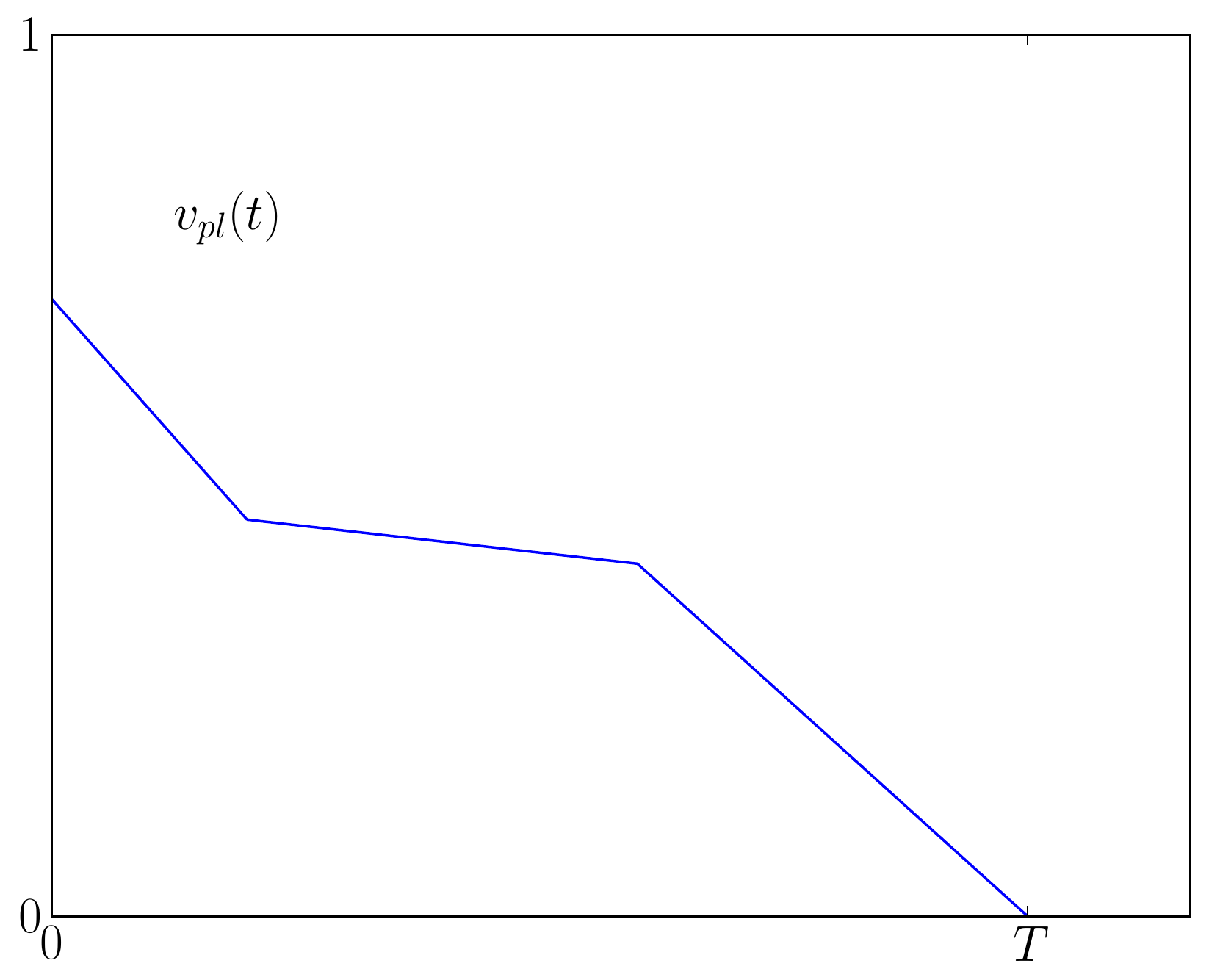}}%
  {A plot of a piecewise linear value decay function.\label{fig:pl}}%
  { The patient has a health score (i.e. internal value) that is
    initially positive and less than 1. As the patient awaits
    treatment, this health score decreases in a continuous piecewise
    linear fashion. This kind of health decay model hsa been used in
    the medical literature, for example in \cite{Sacco_Triage_2005}.}
\end{figure}

Motivated by \cite{Sacco_Triage_2005}, we consider continuous
piecewise linear value decay. Recall that in \cite{Sacco_Triage_2005},
a panel of expert physicians developed a deterministic mechanism for
scoring how patients' health decays over time. As shown earlier in
Figure~\ref{fig:sacco}, the health scores decay continuously and in a
piecewise linear fashion. Maximizing patient health is one way of
optimizing quality of care, so our notion of value is analogous to the
health score from \cite{Sacco_Triage_2005}. Mathematically, we can
write such a value decay function as follows. The interval $[0, T]$ is
divided into $I$ disjoint intervals such that
$$ v_{pl}(t) = \sum_{i=0}^{I-1} (a_i t + b_i)\Ind_{\set{t_i \leq t \leq t_{i+1}}}$$
where $\set{a_i}_{i=0}^{I-1}$ are non-positive constants and
$\set{b_i}_{i=0}^{I-1}$ are non-negative constants. We additionally
require
$$a_i t_{i+1} + b_i = a_{i+1} t_{i+1} + b_{i+t}$$
so that $v_{pl}(\cdot)$ is continuous. For each patient, we randomly
select $\set{a_i}_{i=0}^{I-1}$ and $\set{b_i}_{i=0}^{I-1}$ with an
algorithm detailed in the appendix. An example of a continuous
piecewise linear value decay function is shown in Figure~\ref{fig:pl}.

\begin{figure}[h]
  \FIGURE%
  {
    \begin{minipage}{\textwidth}
      \centering
      \begin{subfigure}{0.48\textwidth}
        \includegraphics[width=\textwidth]{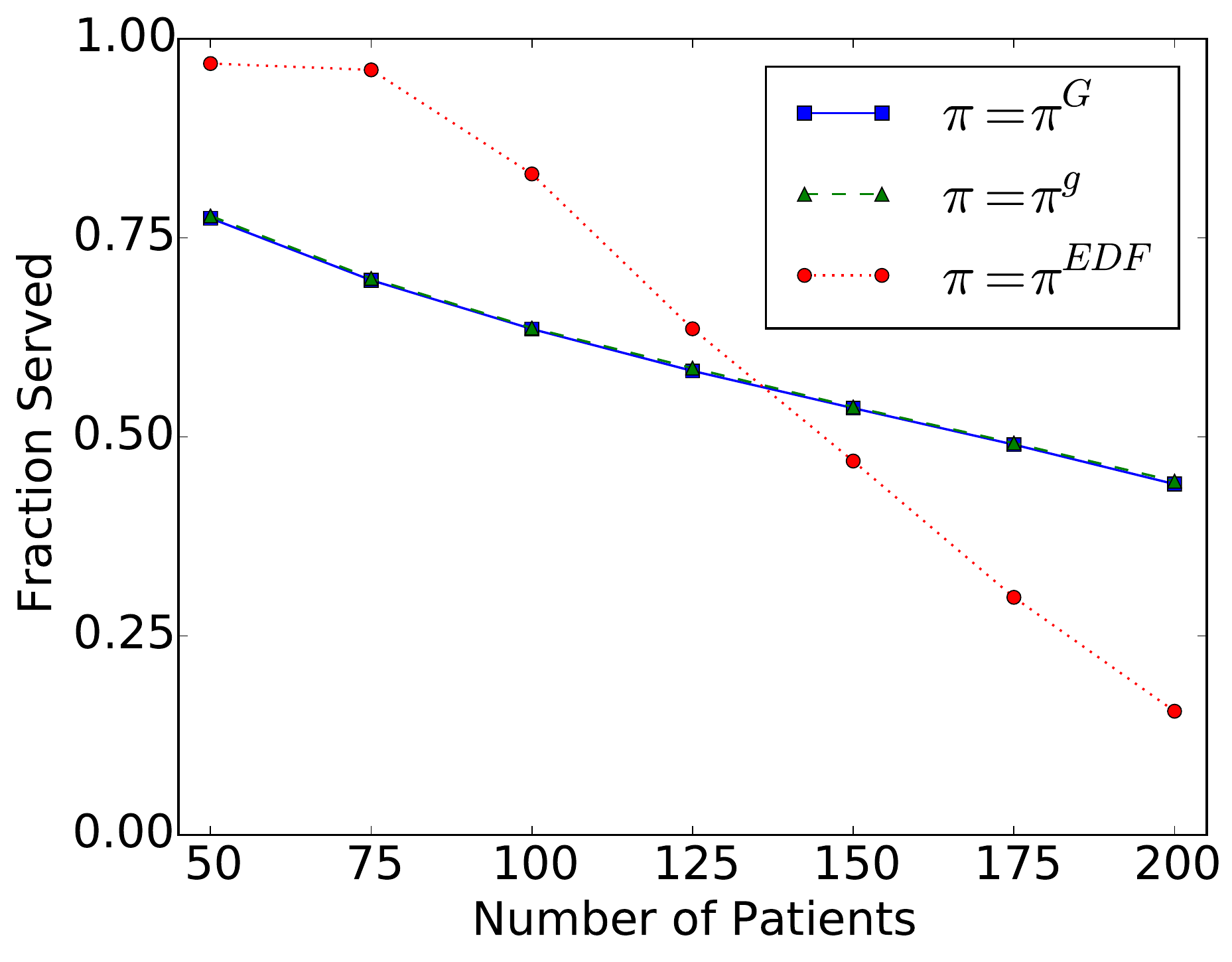}
        \caption{\label{fig:patients_deadline}}
      \end{subfigure}
      \begin{subfigure}{0.48\textwidth}
        \includegraphics[width=\textwidth]{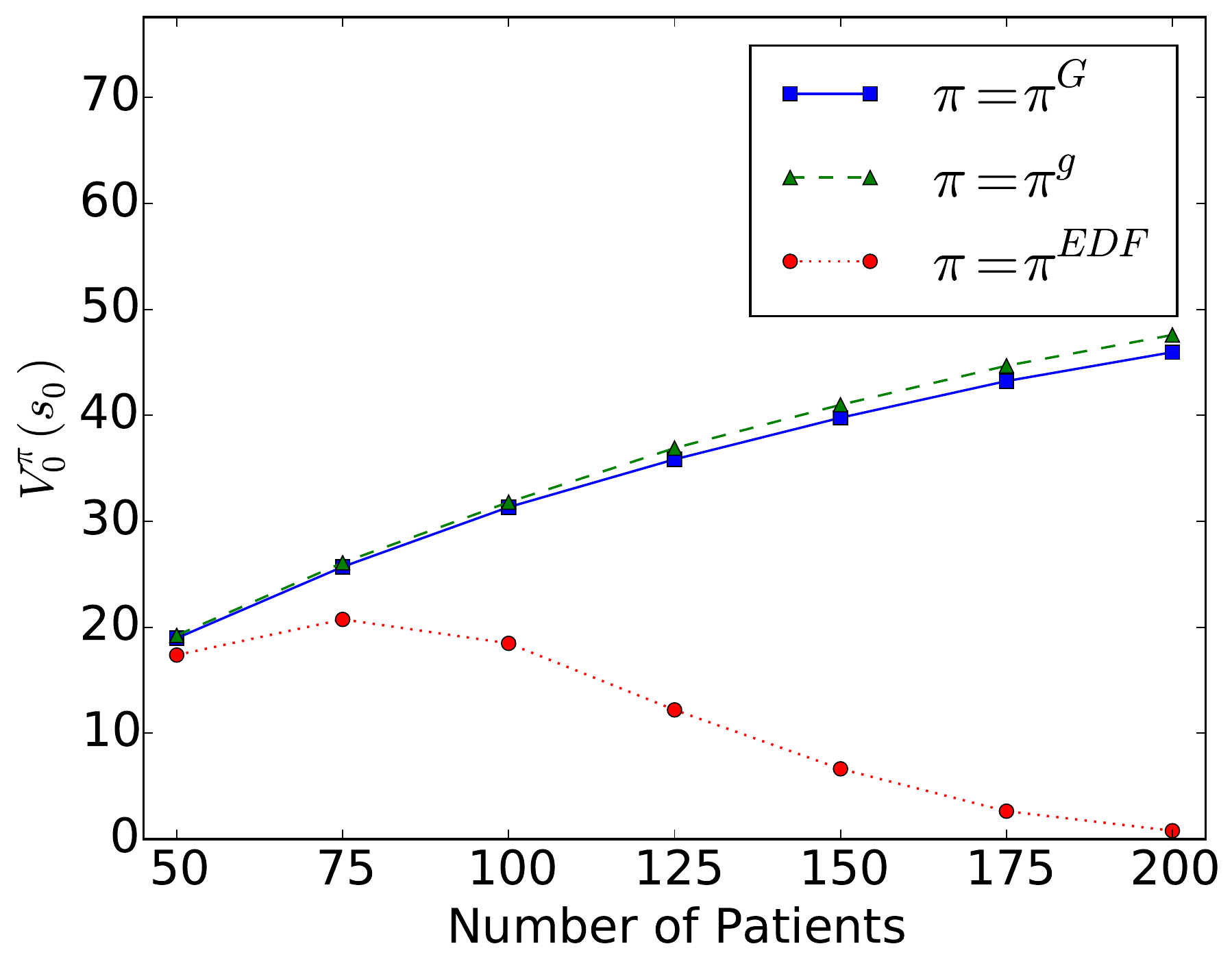}
        \caption{\label{fig:patients_value}}
      \end{subfigure}
    \end{minipage}
  } {Performance of $\pi^G$, $\pi^g$, and $\pi^{EDF}$ in a patient
    scheduling scenario.\label{fig:patients}} {We compare the
    performance of the heuristics when there are $N = 6$ operating
    rooms and $T = 144$ time slots of $\delta = 10$ minute duration
    (i.e. 24 hours divided into 10 minute time slots). The patients
    have piecewise linear value decay and discretized lognormal
    service times. The parameters which characterize each patient are
    chosen randomly. We increase the number of patients $J \in
    \set{50, 75, 100, \hdots, 200}$ and for each $J$ we report the
    average of $1000$ simulations. In addition to reporting
    $V_0^{\pi}(s_0)$, we also report the fraction of patients who are
    served by their final deadline.}
\end{figure}

We vary the number of patients $J \in \set{50, 75, 100, \hdots, 200}$
and compare the performance of $\pi^G$, $\pi^g$, and
$\pi^{EDF}$. Depending on the type of incident, the number of patients
could range from tens (e.g. \cite{Aylwin_London_2007},
\cite{Turegano_Madrid_2008, Turegano_Madrid2_2008}) to hundreds
(e.g. \cite{Cushman_911_2003}), so this is a reasonable range for
$J$. For each $J$, we repeat the simulation $1000$ times and report
the average. In addition to reporting $V^{\pi}_0(s_0)$, we also report
the fraction of patients who are served by their final deadline. The
results are shown in Figure~\ref{fig:patients}.

For $\pi^G$ and $\pi^g$, we see some expected trends: as $J$
increases, both policies increase in value while the fraction of
patients served gradually decreases. As with our other ``large-scale''
numerical experiments, $\pi^g$ performs slightly better than
$\pi^G$. The results for $\pi^{EDF}$ are more nuanced. For $J \leq
75$, the value associated with $\pi^{EDF}$ is less than that of
$\pi^G$ and $\pi^g$, but only slightly.  However, for $J \leq 75$,
$\pi^{EDF}$ serves more than $95\%$ all patients. As a result, for $J
\leq 75$, $\pi^{EDF}$ is better at preventing mortality than $\pi^G$
or $\pi^g$ though potentially at the cost of lower quality of
care. While this is a positive result, we see a decrease in
performance when $J$ increases beyond $100$. For $J \geq 100$, the
value of $\pi^{EDF}$ and the fraction of patients served decreases
substantially. While $\pi^G$ and $\pi^g$ decay in performance
gradually as $J$ increases, $\pi^{EDF}$ exhibits a phase transition
from ``good'' performance to ``poor'' performance. Although the
``critical point'' of this phase transition is difficult to know
\textit{a priori}, this behavior can qualitatively be explained by the
combination of piecewise linear value decay and lognormal service
distributions. Because the patient characteristics are randomly
generated, as $J$ increases, it becomes more likely that patients with
early deadlines have long service times and low health values. This is
due to the fact that the randomly generated piecewise linear value
functions can decay quickly coupled with the fact that lognormal
distributions are heavy-tailed. These patients occupy the operating
rooms and block other patients from being scheduled. This causes many
patients to not be served and for an overall low total value.

These results echo the results that we saw in our previous numerical
experiments. Both $\pi^G$ and $\pi^g$ perform well and are reasonably
robust. In contrast, $\pi^{EDF}$ can perform well sometimes but is not
robust. In general, one will not know \textit{a priori} where
$\pi^{EDF}$ will experience its phase transition from good to bad
performance. Because of this unpredictable behavior, for critical
applications like patient scheduling it is probably best to avoid
$\pi^{EDF}$. On the other hand, $\pi^G$ and $\pi^g$ are both good
choices with $\pi^g$ being slightly better in large-scale scenarios.

\section{Conclusion\label{sec:conclusion}}
We have presented a novel discrete-time model for non-preemptive
scheduling. In this model, jobs have random service times and the
value of each job decays deterministically. The jobs are dynamically
scheduled on identical servers which each have unit service rate. We
formulated the problem in a dynamic programming framework and showed
that while an optimal scheduling policy exists, finding it is
computationally intractable. This leads us to consider three
low-complexity heuristics: a greedy policy, a rate greedy policy and
an earliest-deadline-first (EDF) policy. In addition to providing
performance guarantees (some of which are sharp), we have conducted
extensive numerical experiments to compare the policies.

We have demonstrated that, in general, it is best to use the rate
greedy policy; the greedy policy performs nearly as well as the rate
greedy policy; and, EDF typically does not perform well at
all. However, there are some scenarios in which EDF performs better
than either greedy policy. Specifically, EDF performs better when the
time horizon is long and all of the jobs have a high probability of
completing service in a short amount of time. In all other situations
which we considered, EDF performs poorly. In particular, our
simulations suggest that in patient scheduling scenarios, it is best
to use the rate greedy policy.  We find that it would be reasonable to
use the greedy policy, but EDF likely should be avoided.

Our insights also point us to other research topics of potential
interest. For example, although the heuristics considered in this work
can be applied even when there are job arrivals, our performance
guarantees would no longer be valid. Incorporating job arrivals would
be a slight modeling extension but it would drastically change our
analysis. In particular, our proofs apply backwards induction to the
number of jobs in the system. This requires that the number of jobs in
the system is non-increasing which would clearly be violated if there
were arrivals.  We could also consider a model in which job value
decays stochastically. This would be useful for situations in which
our understanding of the internal job dynamics is imperfect so we only
have a distribution on the value dynamics.  In addition to considering
modeling extensions, within this same model we could consider many
other myopic heuristics which could be relevant to other
applications. As noted above, the proofs can easily be adapted and
extended to other myopic scheduling policies which may be useful for
other applications.

% \ACKNOWLEDGMENT{%
%   Neal Master is funded by the Department of Defense (DoD) through the
%   National Defense Science \& Engineering Graduate (NDSEG) Fellowship
%   Program.  }%

\ACKNOWLEDGMENT{%
  Neal Master is funded by Stanford University through a Stanford
  Graduate Fellowship (SGF) in Science \& Engineering.  }%

\bibliographystyle{ormsv080} % outcomment this and next line in Case 1
\bibliography{myopicScheduling} % if more than one, comma separated

\clearpage
%\ECSwitch
%\ECHead{E-Companion: Proofs}
\begin{APPENDICES}
  \section{Preliminary Mathematical Definitions and Results}
  \subsection{Definitions}
  In addition to the notation in Section~\ref{ssec:dp}, we introduce
  the following notation.

  Given a random variable $X$ the conditional value function for a
  policy $\pi$ is
  $$V_t^\pi(s | X) = \E\left[\left.\sum_{\tau = t}^\infty R_\tau(s_\tau, \pi_\tau(s_\tau))\right| X\right].$$
  If we want to condition on an event $E$, take $X = \Ind_{\set{E}}$
  and let $V_t^\pi(s | E) = V_t^\pi(s | X)\Ind_{\set{X = 1}}$.

  If $s = (b(t), p(t))$ then we let $\bbf(s) = b(t)$ and $\pbf(s) = p(t)$.

  \subsection{Proof of Theorem~\ref{thrm:well-posed}}
  If for some finite $T$ we have $b(T) = (\top, \top, \hdots, \top)$,
  then $p(T) = (0, 0, \hdots, 0)$ and $s_T = (b(T), p(T))$ is a no
  cost/reward trapping state. If we fix any non-idling policy, because
  $\E[\sigma_j] < \infty$ for all $j \in \Jcal$, there is an
  associated finite stopping time $T$ at which $b(T) = (\top, \top,
  \hdots\top)$. Therefore, we have a stochastic shortest path problem
  with destination state $s_T$. The system will reach $s_T$ in a
  finite amount of time and so the given policy is proper. This
  guarantees the existence of an optimal policy which is obtainable
  via policy iteration \cite[Proposition 3.2.2]{Bertsekas_VolII}. \qed

  \subsection{Proof of Theorem~\ref{thrm:NP-hard}}
  Consider a particular instance of the problem in which $\sigma_j$ is
  known with probability 1 and $v_j(t) = c_j \Ind_{\set{t \leq K}}$
  for some fixed constants $\set{c_j}$ and $K$. In this case, the job
  service times are essentially deterministic and the jobs have a
  shared deadline. This is a 0/1 Multiple Knapsack Problem in which
  there are $J$ objects with sizes $\set{\sigma_j}$ and values
  $\set{c_j}$ which need to be placed in $N$ knapsacks each with
  capacity $K$. This particular instance of the problem is NP-hard
  (see \cite{Martello_Knapsack} and the references therein) so the
  problem of computing an optimal policy is NP-hard. \qed

\section{Some Useful Lemmas}
We will now prove a few lemmas which will be useful when proving the
main results of the papers.

Intuitively, if there are more jobs in the system, then there are more
scheduling choices available and an optimal policy will be able to
accrue a greater reward. This intuition is formalized in the following
lemma.
\begin{lemma}[Monotonicity in Jobs]
  \label{lemma:mono_job}
  Consider states $s$ and $s'$ which are related by the three following conditions:
  $$
  \bbf(s')_j \not\in\set{\bot,\top} \implies \bbf(s')_j = \bbf(s)_j;\quad
  \bbf(s')_j = \bot \implies \bbf(s)_j = \bot;\quad
  \pbf(s') = \pbf(s)
  $$
  Then $V_t^*(s) \geq V_t^*(s')$.
\end{lemma}

\proof{Proof.}  Consider a coupling of two systems each starting at
$s$ and $s'$ such that they see the same realizations of service times
$\set{\sigma_j}_{j \in \Jcal}$. Let $\Jcal_{s'} = \set{j \in \Jcal :
  \bbf(s')_j \ne \top}$ and $\Jcal_s = \set{j \in \Jcal : \bbf(s)_j
  \ne \top}$. Note that $\Jcal_{s'} \subseteq \Jcal_s$.

Let $\pi^*$ be an optimal policy. We assume that $\pi^*$ is applied to
the $s'$-system. Consider the following suboptimal policy $\tilde\pi$
which is used for the $s$-system: $\tilde\pi$ takes the same actions
as $\pi^*$ until all the jobs $j \in \Jcal_{s'}$ are completed and
then completes jobs $j \in \Jcal_s \setminus \Jcal_{s'}$ in sequential
order.  Let $T_j$ be the completion time of job $j$ in the $s$-system
when using policy $\tilde\pi$. Let $T_j^*$ be the completion time of
job $j$ in the $s'$-system using the policy $\pi^*$. The coupling
ensures that $T_j = T_j^*$ for all $j \in \Jcal_{s'}$.
\begin{align*}
  V_t^{\tilde\pi}(s | \sigma_1, \hdots, \sigma_J)
  &= \sum_{j \in \Jcal_s} v_j(T_j)
  = \sum_{j \in \Jcal_{s'}} v_j(T_j) + \sum_{j \in \Jcal_s \setminus \Jcal_{s'}} v_j(T_j)
  = \sum_{j \in \Jcal_{s'}} v_j(T_j^*) + \sum_{j \in \Jcal_s \setminus \Jcal_{s'}} v_j(T_j)\\
  &\geq \sum_{j \in \Jcal_{s'}} v_j(T_j^*)
  = V_t^*(s' | \sigma_1, \hdots, \sigma_J)
\end{align*}
Taking the expectation gives us that $V_t^{\tilde\pi}(s) \geq
V_t^*(s')$. The optimality of $V^*$ tells us that $V_t^*(s) \geq
V_t^{\tilde\pi}(s)$ so we can conclude that $V_t^*(s) \geq V_t^*(s')$.
\qed
\endproof

Because the internal value of the jobs decays over time, one would
expect an optimal policy to be non-idling. Our next lemma shows that
it is sufficient to focus our attention on non-idling policies. Note
that because of Lemma~\ref{lemma:nonidling}, we will assume throughout
that any optimal policy is non-idling.
\begin{lemma}[Non-idling]
  \label{lemma:nonidling}
  Suppose that in state $s_t$, there are $M = \abs{\set{n \in \Ncal :
      \pbf(s)_n = 0}}$ free machines and that the number of jobs
  remaining to be processed is $K = \abs{\set{j \in \Jcal : \bbf(s)_j
      = \bot}}$. Then there exists an optimal policy $\pi^*$ such that
  $\pi^*_t(s)$ schedules $\min\set{K, M}$ jobs.
\end{lemma}
\proof{Proof.} Let $A = \pi^*_t(s_t)$. We show that nothing can be
gained by having $\abs{A} < \min\set{K, M}$. Suppose under $\pi^*$
that a processor remains idle even though there is an available
job. Consider another policy $\pi'$ which schedules identically to
$\pi^*$ except it begins processing all jobs on the idling machine one
time slot earlier. Since $v_j(\cdot)$ is non-increasing,
$V_t^{\pi'}(s) \geq V_t^*(s)$. Therefore, $\pi'$ is also optimal and
does not idle. \qed
\endproof

Consider a ``virtual machine'' which is able to complete jobs
instantaneously. Intuitively, such a virtual machine would allow for a
greater total expected reward. Our next lemma justifies this intuition.
\begin{lemma}[Virtual Machine Rewards]
  \label{lemma:vm}
  Fix a state $s$ and a job $i$ such that $\bbf(s)_i = \bot$. Let
  $s'_i  = S'(s, i)$ denote the resulting state if job $i$ were processed
  without occupying a processor. In notation,
  $$
  \bbf(s)_i = \bot,\quad
  \bbf(s'_i)_i = \top,\quad
  \pbf(s) = \pbf(s'_i),\quad
  \bbf(s'_i)_j = \bbf(s)_j\,\forall\, j \in \Jcal\setminus\set{i}.
  $$
  Then $V_t^*(s) \leq \E[v_j(t +
  \sigma_j)] + V_t^*(S'(s, j))$.
\end{lemma}
\proof{Proof.}  Consider two systems starting in states $s$ and $S'(s,
j)$ respectively. We couple the systems so that they see the same
realizations of the services times. We assume that the $s$-system
evolves under an optimal policy $\pi^*$. Let $\set{t_j^*}$ be the
random times are which jobs $j \in \Jcal_s = \set{j \in \Jcal :
  \bbf(s)_j = \bot}$ begin processing. Let $\pi'$ be the policy for
the $s_i'$-system. Let $\set{t_j'}$ be the random times at which job
$j \in \Jcal' = \set{j \in \Jcal : \bbf(s_i')_j = \bot}$. We assume
that $\pi'$ mimics $\pi^*$ as follows. Policy $\pi'$ takes the same
actions as $\pi^*$ for all jobs $j \in \Jcal'$ and when $\pi^*$ would
be processing job $i$, $\pi'$ idles. Therefore, $t_j' = t_j^*$
for all $j \in \Jcal'$:
\begin{align*}
  V_t^*(s)
  &= \E\left[\sum_{j \in \Jcal_s} v_j(t_j^* + \sigma_j)\right]
  = \E\left[\sum_{j \in \Jcal_s\setminus\set{i}} v_j(t_j^* + \sigma_j)\right]
  + \E[v_i(t_i^* + \sigma_i)]\\
  &= \E\left[\sum_{j \in \Jcal'} v_j(t_j' + \sigma_j)\right]
  + \E[v_i(t_i^* + \sigma_i)]
  = V_t^{\pi'}(s_i') + \E[v_i(t_i^* + \sigma_i)]
  \leq V_t^*(s_i') + \E[v_i(t + \sigma_i)]
\end{align*}
The final inequality follows from the optimality of $V_t^*(s_i')$ and
the non-decreasing nature of $v_i(\cdot)$.  \qed
\endproof

\section{Proof of Theorem~\ref{thrm:pi_g}}
The proof of the performance bound for $\pi^g$ is slightly more
complicated than the proofs required for $\pi^G$ and
$\pi^{EDF}$. However, the proofs are structurally similar and so the
later appendices will modify the proof presented here.  We first need
to prove the following lemma which shows how sub-optimal scheduling of
``replica'' jobs can affect the performance of the system.
\begin{lemma}
  \label{lemma:pi_g}
  Consider an augmented set of jobs $\tilde\Jcal = \Jcal \cup
  \set{\tilde{g}, \tilde{i}}$ where $\tilde{g}$ is a replica of some
  job $g$ and $\tilde{i}$ is a replica of some job $i$. By replica, we
  mean that $\sigma_{\tilde{g}} \eqD \sigma_g$ and $\sigma_{\tilde{i}}
  \eqD \sigma_i$ but that $v_{\tilde{g}}(t) = v_{\tilde{i}}(t) =
  0$.

  Consider a state $s_t = s$ at time $t$. Let $\Jcal_s = \set{j \in
    \Jcal : \bbf(s)_j \ne \top}$ and fix $g \in \argmax_{j \in
    \Jcal_s} \frac{\E[v_j(t + \sigma_j)]}{\E[\sigma_j]}$. Assume that
  for some $f \in \Ncal$, $\pbf(s)_f = 0$ (i.e. processor $f$ is free).

  Denote by $s_g$ and $s_i$ two states which are related to state $s$
  in the following manner. The two states are identical to state $s$
  except on machine $f$. In state $s_g$, machine $f$ is occupied by
  the replica $\tilde{g}$ so that $\pbf(s_g)_f = \tilde{g}$. Similarly
  for state $s_i$, $\pbf(s_i)_f = \tilde{i}$. Then
  $$V_t^*(s_i) \leq \left(1 + \frac{\E[\max_{j \in \Jcal_s} \sigma_j]}{\E[\sigma_g]} - \frac{\E[\sigma_i]}{\E[\sigma_g]}\right)\E[v_g(t + \sigma_g)] + V_t^*(s_g).$$
\end{lemma}
Note that because the replica jobs do not generate any rewards, the
augmented system can be optimally controlled by a policy that is
essentially the same as an optimal policy for the original system. If
we take any optimal policy for the original system, we can create an
optimal policy for the augmented system as follows: take the same
actions as the given optimal policy until all jobs in $\Jcal$ have
completed; then complete the replica jobs. Since the replica jobs have
no reward, the optimal value is the same in both systems. As a result
we will abuse notation and use $V_t^*(\cdot)$ as the optimal value
function for all systems in Lemma~\ref{lemma:pi_g}.

\proof{Proof of Lemma~\ref{lemma:pi_g}.}  We begin by coupling the
systems such that they see the same realizations for service
times. Consider a policy $\pi_g'$ for the $s_g$-system which attempts
to mimic the actions taken by an optimal policy $\pi_i^*$ on the
$s_i$-system.  There are two possible cases, $\sigma_{\tilde{i}} \geq
\sigma_{\tilde{g}}$ and $\sigma_{\tilde{i}} < \sigma_{\tilde{g}}$.

\begin{itemize}
\item\underline{Case 1, $\sigma_{\tilde{i}} \geq \sigma_{\tilde{g}}$:}
  After job $\tilde{g}$ has completed, the $\pi_g'$ policy idles on
  machine $f$ until $t+\sigma_{\tilde{i}}$ (time which machine $f$ is
  free in the $s_i$-system).  At this point, the $s_g$-system is
  ``synced'' with the $s_i$-system and it proceeds with executing the
  optimal policy for the $s_i$ system, $\pi_i^*$.

  If $T_j^*(s_i)$ is the completion time of job $j$ in the
  $s_i$-system under optimal policy $\pi_i^*$, and $T_j$ is the
  completion time of job $j$ in the $s_g$-system under the $\pi_g'$,
  then $T_j = T_j^*(s_i)$ for all $j \in \Jcal_s$.
  \begin{align}\label{eqn:gr1}
    \begin{split}
      V_t^*(s_i|\sigma_{\tilde{i}} \geq \sigma_{\tilde{g}})
      &= \E\left[\left.\sum_{j\in \Jcal_s} v_j(T_j^*(s_i))
        \right|\sigma_{\tilde{i}} \geq \sigma_{\tilde{g}}\right]
      = \E\left[\left.\sum_{j\in \Jcal_s} v_j(T_j)
        \right|\sigma_{\tilde{i}} \geq \sigma_{\tilde{g}}\right]
      = V_t^{\pi_g'}(s_g|\sigma_{\tilde{i}} \geq \sigma_{\tilde{g}}) \\
      &\leq V_t^*(s_g|\sigma_{\tilde{i}} \geq \sigma_{\tilde{g}})\\
      &\leq V_t^*(s_g|\sigma_{\tilde{i}} \geq \sigma_{\tilde{g}})
      + \left(1 + \frac{\E[\max_{j\in\Jcal_s}\sigma_j]}{\E[\sigma_g]}
        -\frac{\E[\sigma_i]}{\E[\sigma_g]}\right)\E[v_g(t+\sigma_g)]
      \end{split}
  \end{align}
  The inequalities come from the construction of $\pi_g'$ and the fact
  that the added term is non-negative.

\item\underline{Case 2, $\sigma_{\tilde{i}}< \sigma_{\tilde{g}}$:}
  In this case, $\pi_g'$ cannot exactly mimic $\pi_i^*$ policy because
  machine $f$ will continue to be busy after $\tilde{i}$ completes in
  the $s_i$-system. While machine $f$ is processing $\tilde{g}$ in the
  $s_g$-system, machine $f$ will process jobs $\Jcal_{sim}$ in the
  $s_i$-system. The $\pi_g'$ policy will ``simulate'' the processing
  of the jobs in $\Jcal_{sim}$ while they are actually being processed
  in the $s_i$-system. After job $\tilde{g}$ is completed in the
  $s_g$-system, $\pi_g'$ will continue to mimic $\pi_i^*$ as if the
  simulated jobs were actually completed. When $\pi_i^*$ has completed
  all jobs in the $s_i$-system, $\pi_g'$ will then complete jobs in
  $\Jcal_{sim}$ in the $s_g$-system in some arbitrary
  order.

  If $T_j^*(s_i)$ is the completion time of job $j$ in the
  $s_i$-system under optimal policy, $\pi_i^*$ and $T_j$ is the
  completion time of job $j$ in the $s_g$-system under the $\pi_g'$
  policy, then $T_j = T_j^*(s_i)$ for all $j \in \Jcal \setminus
  \Jcal_{sim}$.

  \begin{align}\label{eqn:gr_help}
    \begin{split}
      V_t^*(s_i|\sigma_{\tilde{i}} < \sigma_{\tilde{g}})
      &= \E\left[\left.
          \sum_{j \in \Jcal_s} v_j(T_j^*(s_i))
        \right| \sigma_{\tilde{i}} < \sigma_{\tilde{g}}\right]\\
      &= \E\left[\left.
          \sum_{j \in \Jcal_{sim}} v_j(T_j^*(s_i)) + \sum_{j \not \in \Jcal_{sim}} v_j(T_j^*(s_i))
        \right| \sigma_{\tilde{i}} < \sigma_{\tilde{g}}\right]\\
      &\leq \E\left[\left.
          \sum_{j \in \Jcal_{sim}} v_j(T_j^*(s_i)) +
          \sum_{j \not \in \Jcal_{sim}} v_j(T_j^*(s_i)) + \sum_{j \in \Jcal_{sim}} v_j(T_j)
        \right| \sigma_{\tilde{i}} < \sigma_{\tilde{g}}\right]\\
      &= \E\left[\left.
          \sum_{j \in \Jcal_{sim}} v_j(T_j^*(s_i))
        \right| \sigma_{\tilde{i}} < \sigma_{\tilde{g}}\right]
      + V_t^{\pi_g'}(s_g|\sigma_{\tilde{i}} < \sigma_{\tilde{g}})\\
      &\leq \E\left[\left.
          \sum_{j \in \Jcal_{sim}} v_j(T_j^*(s_i))
        \right| \sigma_{\tilde{i}} < \sigma_{\tilde{g}}\right]
      + V_t^*(s_g|\sigma_{\tilde{i}} < \sigma_{\tilde{g}})
    \end{split}
  \end{align}
  The first equality follows by definition. The second equality
  follows because $\Jcal_s = \Jcal_{sim} \cup \Jcal_{sim}^c$. The
  inequality comes from the non-negativity of $v_j(\cdot)$. The final
  two equalities follow from the construction of $\pi_g'$.

  Continuing, note that at the earliest, job $j$ can be completed at
  $t + \sigma_j$ so $T_j^*(s_i) \geq t + \sigma_j$. Since $v_j(\cdot)$
  is non-decreasing, we have the following:
  \begin{align}\label{eqn:needsMod}
    \begin{split}
      V_t^*(s_i|\sigma_{\tilde{i}} < \sigma_{\tilde{g}})
      &\leq \E\left[\left.
          \sum_{j \in \Jcal_{sim}} v_j(T_j^*(s_i))
        \right| \sigma_{\tilde{i}} < \sigma_{\tilde{g}}\right]
      + V_t^*(s_g|\sigma_{\tilde{i}} < \sigma_{\tilde{g}})\\
      &\leq \E\left[\left.
          \sum_{j \in \Jcal_{sim}} v_j(t + \sigma_j)
        \right| \sigma_{\tilde{i}} < \sigma_{\tilde{g}}\right]
      + V_t^*(s_g|\sigma_{\tilde{i}} < \sigma_{\tilde{g}})
    \end{split}
  \end{align}
  Now focus on the first term. Since $\Jcal_{sim}$ depends on the
  realizations of $\sigma_{\tilde{i}}$ and $\sigma_{\tilde{g}}$, it is
  a random set upon which we can condition.
  \begin{align*}
    \E\left[\left.
        \sum_{j \in \Jcal_{sim}} v_j(t + \sigma_j)
      \right| \sigma_{\tilde{i}} < \sigma_{\tilde{g}}, \Jcal_{sim}\right]
    &= \E\left[\left.
        \sum_{j \in \Jcal_{sim}} \E[\sigma_j]\frac{v_j(t + \sigma_j)}{\E[\sigma_j]}
      \right| \sigma_{\tilde{i}} < \sigma_{\tilde{g}}, \Jcal_{sim}\right] \\
    &= \sum_{j \in \Jcal_{sim}} \E[\sigma_j]
    \frac{\E[v_j(t + \sigma_j)| \sigma_{\tilde{i}} < \sigma_{\tilde{g}}, \Jcal_{sim}]}
    {\E[\sigma_j]}\\
    &\leq \sum_{j \in \Jcal_{sim}} \E[\sigma_j]\frac{\E[v_g(t + \sigma_g)]}{\E[\sigma_g]}
    = \frac{\E[v_g(t + \sigma_g)]}{\E[\sigma_g]}\sum_{j \in \Jcal_{sim}} \E[\sigma_j]
  \end{align*}

  The first equality is valid because $\sigma_j \geq 1$ and the second
  equality is an application of the linearity of conditional
  expectation. The final inequality is due to the fact that the index
  $g$ is chosen independently of $\sigma_{\tilde{i}}$,
  $\sigma_{\tilde{g}}$, and $\Jcal_{sim}$.

  The maximum amount of time machine $f$ will idle in the $s_g$-system
  will be $\sigma_{\tilde{g}} - \sigma_{\tilde{i}} +
  \sigma_{max}$. Indeed, $\pi_g'$ will need to simulate jobs for at
  least $\sigma_{\tilde{g}} - \sigma_{\tilde{i}}$ and one job may
  begin simulation just before $t + \sigma_{\tilde{g}}$. Applying this
  bound and taking the expectation of $\Jcal_{sim}$ gives the
  following:
  \begin{align*}
    \E\left[\left.
        \sum_{j \in \Jcal_{sim}} v_j(t + \sigma_j)
      \right| \sigma_{\tilde{i}} < \sigma_{\tilde{g}}, \right]
    &\leq \frac{\E[v_g(t + \sigma_j)]}{\E[\sigma_g]}
    \E\left[\left.\sum_{j \in \Jcal_{sim}} \E[\sigma_j]
      \right| \sigma_{\tilde{i}} < \sigma_{\tilde{g}}\right]\\
    &\leq \frac{\E[v_g(t + \sigma_j)]}{\E[\sigma_g]}
    \E[\sigma_{\tilde{g}} - \sigma_{\tilde{i}} + \sigma_{max}
    | \sigma_{\tilde{i}} < \sigma_{\tilde{g}}]
  \end{align*}
  So we have that
  \begin{equation}
    V_t^*(s_i | \sigma_{\tilde{i}} < \sigma_{\tilde{g}})
    \leq
    \frac{\E[v_g(t + \sigma_j)]}{\E[\sigma_g]}
    \E[\sigma_{\tilde{g}} - \sigma_{\tilde{i}} + \sigma_{max}
    | \sigma_{\tilde{i}} < \sigma_{\tilde{g}}]
    + V_t^*(s_g | \sigma_{\tilde{i}} < \sigma_{\tilde{g}}).
    \label{eqn:gr2}
  \end{equation}
\end{itemize}

Now combine (\ref{eqn:gr1}) and (\ref{eqn:gr2}) and take the
expectation over $\sigma_{\tilde{i}} \geq \sigma_{\tilde{g}}$ and
$\sigma_{\tilde{i}} < \sigma_{\tilde{g}}$. Noting that
$\sigma_{\tilde{i}} \eqD \sigma_i$ and $\sigma_{\tilde{g}} \eqD
\sigma_g$ gives us the result:
\begin{align*}
    V_t^*(s_i)
    \leq
    \frac{\E[v_g(t+\sigma_g)]}{\E[\sigma_g]}(\E[\sigma_g]-\E[\sigma_i]+\E[\sigma_{\max}])
    + V_t^*(s_g)
    =
    \E[v_g(t+\sigma_g)]\left(
      1 + \frac{\E[\sigma_{max}]}{\E[\sigma_g]} - \frac{\E[\sigma_i]}{\E[\sigma_g]}\right)
    + V_t^*(s_g)
\end{align*}
\qed
\endproof

Now we can prove the performance guarantee in
Theorem~\ref{thrm:pi_g}. One of the key ideas of the proof is as
follows: we will add replica jobs as in Lemma~\ref{lemma:pi_g}, apply
the monotonicity result in Lemma~\ref{lemma:mono_job}, and make use of
the virtual machines in Lemma~\ref{lemma:vm} to complete the
replicas. Because the replica jobs have no value, ``adding and
subtracting'' these replicas does not impact the total value of
system.

\proof{Proof of Theorem~\ref{thrm:pi_g}} The proof proceeds by
induction on the number of jobs remaining to be processed, $\sum_{j\in
  \Jcal} \Ind_{\{b_j(t) = \bot\}}$.  The claim is trivially true if
there is only one job remaining to be processed because $\pi^g$ and
any (non-idling) $\pi^*$ will coincide. Now consider a state $s$ such
that $\sum_j \Ind_{\{\bbf(s)_j = \bot\}} = K$, and assume that the
claim is true for all states $s'$ with $\sum_j \Ind_{\{\bbf(s')_j =
  \bot\}} < K$.

Now if $\pi_t^*(s) = \pi_t^g(s)$ the then the next state encountered
and rewards generated in both systems are identically distributed so
that the induction hypothesis immediately yields the result for state
$s$.

Consider the case where $\pi_t^*(s) \neq \pi_t^g(s)$. Let $A_*$ and
$A_g$ denote the optimal and myopic scheduling policy, respectively,
given state $s$ in time slot $t$. Denote by $\Jcal_*$ and $\Jcal_g$
the corresponding sets of jobs processed by the optimal and myopic
policies in state $s$ at time $t$. We suppress the dependence on $s$
and $t$ for notational compactness.  Recall that by
Lemma~\ref{lemma:nonidling}, $|\Jcal_*| = |\Jcal_g|$. Furthermore, the
free processors which are being assigned jobs define a bijection
between $\Jcal_*$ and $\Jcal_g$. If we take $i \in \Jcal_*$ then $g(i)
\in \Jcal_g$ is corresponding job.  Similarly, if we take $g \in
\Jcal_g$ then $i(g) \in \Jcal_*$ is the corresponding job. The myopic
policy will select the $\abs{\Jcal_g}$ jobs with the largest reward
rate. Since the processors are identical, we can therefore assume that
jobs are matched to processors in a way so that $\frac{\E[v_i(t +
  \sigma_i)]}{\E[\sigma_i]} \leq \frac{\E[v_{g(i)}(t +
  \sigma_{g(i)})]}{\E[\sigma_{g(i)}]}$.

Taking definitions from before, we define $\tilde{S}(s,A)$ as the
random next state encountered given that we start in state $s$ and
action $A$ is taken.  Also, $S'(s,i)$ is identical to state $s$ but
with job $i$ is completed: $\bbf(S'(s, i))_i = \top$.

Given a scheduling action $A$ and state $s$ at time $t$, we define the
augmented state $\hat{s} = \hat{S}(s, A)$ as follows. Let $\Jcal_A$
denote the jobs to be scheduled by $A$ and $\tilde\Jcal_A$ to be
replicas of these jobs. Recall that replica jobs have the same service
requirements but no reward: if $j \in \Jcal_A$ and $\tilde{j} \in
\tilde\Jcal_A$ is the replica, then $\sigma_{\tilde{j}} \eqD
\sigma_{j}$ and $v_{\tilde{j}}(\cdot) = 0$. Then $\hat{s}$ is given by
scheduling the replica jobs instead of the original jobs. In notation,
$\bbf(\hat{s})_j = \bbf(s)_j$ for $j \in \Jcal_A$,
$\bbf(\hat{s})_{\tilde{j}} = t$ for $\tilde{j} \in \tilde\Jcal_A$,
$\pbf(\hat{s})_n = \tilde{j}$ for $(j, n) \in A$, and $\pbf(\hat{s})_n
= \pbf(s)_n$ otherwise.

Using this notation, we have the following:
\begin{align}
  \label{eq:th1}
  \begin{split}
    V_t^*(s)
    &= \sum_{j\in\Jcal_*}\E[v_j(t+\sigma_j)] + \E[V_t^*(\tilde{S}(s,A_*))]
    = \sum_{j\in\Jcal_*}\E[\sigma_j]\frac{\E[v_j(t+\sigma_j)]}{\E[\sigma_j]}
    + \E[V_t^*(\tilde{S}(s,A_*))]\\
    &\leq \sum_{i\in\Jcal_*}\E[\sigma_i]\frac{\E[v_{g(i)}(t+\sigma_{g(i)})]}{\E[\sigma_{g(i)}]}
    + \E[V_t^*(\tilde{S}(s,A_*))]\\
    &\leq
    \sum_{i\in\Jcal_*}\E[\sigma_i]\frac{\E[v_{g(i)}(t+\sigma_{g(i)})]}{\E[\sigma_{g(i)}]}
    + \E[V_t^*(\hat{S}(s,A_*))]\\
  \end{split}
\end{align}

The first inequality comes from the definition of the myopic policy;
the reward rate for myopic jobs is higher than for the optimal jobs.
If we consider the replica jobs in $\hat{S}(s, A_*)$ as being
completed in $\tilde{S}(s, A_*)$ (this is consistent because the
replicas have no reward), the second inequality comes from the
monotonicity property proven in Lemma \ref{lemma:mono_job}.

Consider the second term in (\ref{eq:th1}). Because we have a
bijection between jobs scheduled by $\pi^*$ and jobs scheduled by
$\pi^g$, we can apply Lemma \ref{lemma:pi_g} to each job-processor to
conclude the following:
$$
\E[V_t^*(\hat{S}(s, A_*))] \leq \E[V_t^*(\hat{S}(s, A_g))] + \sum_{g
  \in \Jcal_g}\E[v_g(t + \sigma_g)]\left(1 -
  \frac{\E[\sigma_{i(g)}]}{\E[\sigma_g]} +
  \frac{\E[\sigma_{max}]}{\E[\sigma_g]}\right)
$$
Substituting this into (\ref{eq:th1}) gives us the following:
\begin{align}
  \label{eq:th2}
  V_t^*(s) &\leq \E[V_t^*(\hat{S}(s, A_g))] +
  \sum_{g \in \Jcal_g}\E[v_g(t + \sigma_g)]
  \left(1 + \frac{\E[\sigma_{max}]}{\E[\sigma_g]}\right)
\end{align}

Now consider the upper bound in (\ref{eq:th2}):
\begin{align}
  \label{eq:th3}
  \begin{split}
    \E[V_t^*(\hat{S}(s, A_g))]
    + \sum_{g \in \Jcal_g}\E[v_g(t + \sigma_g)] \left(1 +
      \frac{\E[\sigma_{max}]}{\E[\sigma_g]}\right)
    &\leq
    \E[V_t^*(\tilde{S}(s, A_g))]
    + \sum_{g \in \Jcal_g}\E[v_g(t + \sigma_g)] \left(2 +
      \frac{\E[\sigma_{max}]}{\E[\sigma_g]}\right)\\
    &\leq
    \E[V_t^*(\tilde{S}(s, A_g))]
    + \sum_{g \in \Jcal_g}\E[v_g(t + \sigma_g)] (2 + \Delta)\\
    &\leq
    (2 + \Delta)\E[V_t^g(\tilde{S}(s, A_g))]
    + \sum_{g \in \Jcal_g}\E[v_g(t + \sigma_g)] (2 + \Delta)\\
    &= (2 + \Delta)V_t^g(s)
    \end{split}
  \end{align}
  The first inequality comes from applying Lemma~\ref{lemma:vm} to
  each processor being scheduled. The second inequality comes from the
  definition of $\Delta$. The third inequality comes from the
  induction hypothesis. The final equality comes from the Bellman
  recursion corresponding to $\pi^g$. We can conclude that $V_t^*(s)
  \leq (2+\Delta)V_t^g(s)$.  \qed
\endproof

\section{Proof of Theorem~\ref{thrm:pi_G}}
To prove Theorem~\ref{thrm:pi_G}, we will need to slightly modify the
results from above. First we will modify Lemma~\ref{lemma:pi_g} so
that the replica jobs correspond to those scheduled by $\pi^G$ rather
than $\pi^g$.

\begin{lemma}
  \label{lemma:pi_G}
  Modify the conditions of Lemma~\ref{lemma:pi_g} so that $g \in
  \argmax_{j \in \Jcal_s} \E[w_j(t + \sigma_j)]$.  Then we have the
  following inequality:
  $$
  V_t^*(s_i) \leq
  \E[v_g(t + \sigma_g)]\E\left[\frac{\sigma_{max} - \sigma_i +
      \sigma_g}{\sigma_{min}}\right]
  + V_t^*(s_g)
  $$
\end{lemma}
\proof{Proof.}
  As in (\ref{eqn:gr1}), $V_t^*(s_i | \sigma_{\tilde{i}} \geq
  \sigma_{\tilde{g}}) \leq V_t^*(s_g | \sigma_{\tilde{i}} \geq
  \sigma_{\tilde{g}})$. Therefore,
  \begin{equation}
    V_t^*(s_i | \sigma_{\tilde{i}} \geq \sigma_{\tilde{g}}) \leq
    \E[v_g(t + \sigma_g)]\E\left[\frac{\sigma_{max} - \sigma_i + \sigma_g}{\sigma_{min}}\right]
    + V_t^*(s_g | \sigma_{\tilde{i}} \geq \sigma_{\tilde{g}})
  \end{equation}

  For the case that $\sigma_{\tilde{i}} < \sigma_{\tilde{g}}$, define
  $\pi_g'$ and $\pi_i^*$ as in Lemma \ref{lemma:pi_g}. As in
  (\ref{eqn:needsMod}),
  $$
  V_t^*(s_i | \sigma_{\tilde{i}} < \sigma_{\tilde{g}})
  \leq \E\left[\left.
      \sum_{j \in \Jcal_{sim}} v_j(t + \sigma_j) \right| \sigma_{\tilde{i}} < \sigma_{\tilde{g}}
  \right]
  + V_t^*(s_g | \sigma_{\tilde{i}} < \sigma_{\tilde{g}}).
  $$
  Now we bound the first term.
  \begin{align*}
    \E\left[\left.
        \sum_{j \in \Jcal_{sim}} v_j(t + \sigma_j) \right| \sigma_{\tilde{i}} < \sigma_{\tilde{g}}
    \right]
    &\leq \E\left[\left.
        \sum_{j \in \Jcal_{sim}} v_g(t + \sigma_g) \right| \sigma_{\tilde{i}} < \sigma_{\tilde{g}}
    \right]\\
    &\leq \E[v_g(t + \sigma_g) | \sigma_{\tilde{i}} < \sigma_{\tilde{g}}]
    \E[\abs{\Jcal_{sim}}| \sigma_{\tilde{i}} < \sigma_{\tilde{g}}]\\
    &\leq \E[v_g(t + \sigma_g)]\E[\abs{\Jcal_{sim}}| \sigma_{\tilde{i}} < \sigma_{\tilde{g}}]\\
    &\leq \E[v_g(t + \sigma_g)]
    \E\left[\left.\frac{1}{\sigma_{min}}\sum_{j \in \Jcal_{sim}} \sigma_j\right|
      \sigma_{\tilde{i}} < \sigma_{\tilde{g}}\right]\\
    &\leq \E[v_g(t + \sigma_g)]
    \E\left[\left.
        \frac{\sigma_{\tilde{g}} - \sigma_{\tilde{i}} + \sigma_{max}}{\sigma_{min}}\right|
      \sigma_{\tilde{i}} < \sigma_{\tilde{g}}\right]\\
    &= \E[v_g(t + \sigma_g)]
    \E\left[\left.
        \frac{\sigma_g - \sigma_i + \sigma_{max}}{\sigma_{min}}\right|
      \sigma_{\tilde{i}} < \sigma_{\tilde{g}}\right]
  \end{align*}
  The first inequality follows from the definition of job $g$. The
  second inequality follows from an elementary application of Wald's
  identity. The third inequality follows because job $g$ is chosen
  independently from $\sigma_{\tilde{i}}$ and
  $\sigma_{\tilde{g}}$. The fourth inequality follows because the
  number of simulated jobs is upper bounded by the simulation duration
  divided by the smallest amount of time it takes to simulate a
  job. The fifth inequality follows from the same reasoning in Lemma
  \ref{lemma:pi_g}. The final equality follows because
  $\sigma_{\tilde{i}} \eqD \sigma_i$, $\sigma_{\tilde{g}} \eqD
  \sigma_g$, and the service times are independent. Therefore,
  $$
  V_t^*(s_i | \sigma_{\tilde{i}} < \sigma_{\tilde{g}})
  \leq \E[v_g(t + \sigma_g)]
  \E\left[\left.
      \frac{\sigma_g - \sigma_i + \sigma_{max}}{\sigma_{min}}\right|
    \sigma_{\tilde{i}} < \sigma_{\tilde{g}}\right]
  + V_t^*(s_g | \sigma_{\tilde{i}} < \sigma_{\tilde{g}}).
  $$
  Combining the two cases and taking the expectation gives us the
  result.
\qed
\endproof

The proof of Theorem~\ref{thrm:pi_G} is now similar to the proof of
Theorem~\ref{thrm:pi_g}: we will add zero-value replica jobs as in
Lemma~\ref{lemma:pi_g}, apply the monotonicity result in
Lemma~\ref{lemma:mono_job}, and make use of the virtual machines in
Lemma~\ref{lemma:vm} to complete the replicas.

\proof{Proof of Theorem~\ref{thrm:pi_G}} As in Theorem
\ref{thrm:pi_g}, we proceed by induction on the number of jobs
remaining to be processed. When there is one job left, $\pi^*$ and
$\pi^G$ will coincide and the bound holds.

Now we modify the notation in Theorem \ref{thrm:pi_g} so that
$\Jcal_g$ refers to the jobs chosen by $\pi^G$ rather than $\pi^g$.
\begin{align*}
  V_t^*(s)
  &= \sum_{j \in \Jcal_*} \E[v_j(t + \sigma_j)] + \E[V_t^*(\tilde{S}(s, A_*))]\\
  &\leq \sum_{j \in \Jcal_*} \E[v_j(t + \sigma_j)] + \E[V_t^*(\hat{S}(s, A_*))]\\
  &\leq \sum_{g \in \Jcal_*} \E[v_g(t + \sigma_g)] + \E[V_t^*(\hat{S}(s, A_*))]
\end{align*}
The equality comes from the Bellman recursion. The first inequality
comes from Lemma \ref{lemma:mono_job} and the second inequality comes
from the definition of $\pi^G$.

In Theorem \ref{thrm:pi_g}, we applied Lemma~\ref{lemma:pi_g} to each
processor. Here, we apply Lemma \ref{lemma:pi_G}. As before, let
$i(g)$ denote the job index that $\pi^*$ would schedule instead of job
$g$ that $\pi^G$ is scheduling.
\begin{equation*}
  \E[V_t^*(\hat{S}(s, A_*))] \leq
  \sum_{g \in \Jcal_g}
  \left(\E[v_g(t + \sigma_g)]
    \E\left[\frac{\sigma_{max} - \sigma_{i(g)} + \sigma_g}{\sigma_{min}}\right]\right)
  + \E[V_t^*(\hat{S}(s, A_g))]
\end{equation*}
The fact that $\sigma_{i(g)} / \sigma_{min} \geq 1$ gives us that
$\E[v_g(t + \sigma_g)] \leq \E[v_g(t +
\sigma_g)]\E[\sigma_{i(g)}/\sigma_{min}]$. Therefore, combining the
inequalities gives us that
\begin{equation*}
  V_t^*(s) \leq
  \sum_{g \in \Jcal_g} \E[v_g(t + \sigma_g)]
  \E\left[\frac{\sigma_{max} + \sigma_g}{\sigma_{min}}\right]
  + \E[V_t^*(\hat{S}(s, A_g))].
\end{equation*}
Applying Lemma \ref{lemma:vm} gives us that
\begin{equation*}
  V_t^*(s) \leq
  \sum_{g \in \Jcal_g} \E[v_g(t + \sigma_g)]\left(1 +
    \E\left[\frac{\sigma_{max} + \sigma_g}{\sigma_{min}}\right]\right)
  + \E[V_t^*(\tilde{S}(s, A_g))].
\end{equation*}
Now we need to apply some algebraic manipulations:
\begin{align}
  V_t^*(s) &\leq
  \sum_{g \in \Jcal_g} \E[v_g(t + \sigma_g)]\left(1 +
    \E\left[\frac{\sigma_{max} + \sigma_g}{\sigma_{min}}\right]\right)
  + \E[V_t^*(\tilde{S}(s, A_g))]\\
  &\leq
  (1 + 2\E[\sigma_{max}/\sigma_{min}])\sum_{g \in \Jcal_g} \E[v_g(t + \sigma_g)]
  + \E[V_t^*(\tilde{S}(s, A_g))]\\
  &\leq
  (1 + 2\E[\sigma_{max}/\sigma_{min}])\sum_{g \in \Jcal_g} \E[v_g(t + \sigma_g)]
  + \E[(1 + 2\E[\sigma_{max}/\sigma_{min}])V_t^*(\tilde{S}(s, A_g))]\\
  &\leq
  (1 + 2\E[\sigma_{max}/\sigma_{min}])\left(\sum_{g \in \Jcal_g} \E[v_g(t + \sigma_g)]
    + V_t^*(\tilde{S}(s, A_g))]\right)\\
  &= (1 + 2\E[\sigma_{max}/\sigma_{min}])V_t^G(s)
  \end{align}
  The second inequality hold because $\sigma_g \leq \sigma_{max}$. We
  then apply the induction hypothesis and the Bellman recursion to
  achieve the result.
\qed
\endproof

\section{Proof of Proposition~\ref{prop:2approx}}
In this section, we consider $\pi^g$ and $\pi^G$ in the case of IID
service times. We will again modify the proofs of
Lemma~\ref{lemma:pi_g} and Theorem~\ref{thrm:pi_g}. 

\begin{lemma}
  \label{lemma:2approx}
  Consider the notation in Lemma~\ref{lemma:pi_g} but let $g$ and $i$
  be arbitrary job indices. If $\sigma_1 \eqD \sigma_2 \eqD \cdots
  \eqD \sigma_J$, then $V_t^*(s_i) = V_t^*(s_g)$.
\end{lemma}
\proof{Proof.}  Since $\sigma_{\tilde{i}} \eqD \sigma_{\tilde{g}}$ and
$v_{\tilde{i}}(\cdot) = v_{\tilde{g}}(\cdot) = 0$, the $s_i$-system
and the $s_g$-system are stochastically equivalent. Therefore,
$V_t^*(s_i) = V_t^*(s_g)$. \qed
\endproof

\proof{Proof of Proposition~\ref{prop:2approx}} Under this scenario, Lemma
\ref{lemma:pi_g} can be replaced by Lemma \ref{lemma:2approx} in the
proof of Theorem~\ref{thrm:pi_g}.  Hence, $\E[V_t^*(\hat{S}(s,A_*))] =
\E[V_t^*(\hat{S}(s,A_g))]$. Instead of replicating the entire proof
here, we examine how (\ref{eq:th1}), (\ref{eq:th2}), and
(\ref{eq:th3}) change.

The only difference for (\ref{eq:th1}) is that $\E[\sigma_j] =
\E[\sigma_i]$ for $i,j$ which allows for a slight simplification.
\begin{align}
  \begin{split}
    V_t^*(s)
    &= \sum_{j\in\Jcal_*}\E[v_j(t+\sigma_j)] + \E[V_t^*(\tilde{S}(s,A_*))]
    \leq \sum_{g\in \Jcal_g}\frac{\E[\sigma_{i(g)}]}{\E[\sigma_g]}\E[v_g(t+\sigma_g)]
    + \E[V_t^*(\tilde{S}(s,A_*))]\\
    &\leq \sum_{g\in\Jcal_g}\E[v_g(t+\sigma_g)] +  \E[V_t^*(\hat{S}(s,A_*))]
  \end{split}
\end{align}
Now, with improvement to Lemma \ref{lemma:pi_g} in Lemma
\ref{lemma:2approx}, (\ref{eq:th2}) is reduced significantly
\begin{align}
  \sum_{g \in \Jcal_g}\E[v_g(t+\sigma_g)] + \E[V_t^*(\hat{S}(s,A_*))]
  =  \sum_{g\in\Jcal_g}\E[v_g(t+\sigma_g)] +  \E[V_t^*(\hat{S}(s,A_g))]
\end{align}
Finally, utilizing Lemma \ref{lemma:vm} and completing/generating
rewards for the myopic jobs gives:
\begin{align}
  \begin{split}
    \sum_{g\in\Jcal_g}\E[v_g(t+\sigma_g)] +  \E[V_t^*(\hat{S}(s,A_g))]
    &\leq 2\sum_{g\in\Jcal_g}\E[v_g(t+\sigma_g)] + \E[V_t^*(\tilde{S}(s,A_g))]\\
    &\leq 2\sum_{g\in\Jcal_g}\E[v_g(t+\sigma_g)] + 2\E[V_g^*(\tilde{S}(s,A_g))] \\
    &=2 V_t^g(s)
  \end{split}
\end{align}
\qed
\endproof

\section{Proof of Theorem~\ref{thrm:pi_EDF}}
The performance guarantee for $\pi^{EDF}$ only holds for IID service
times. Since we already have Lemma~\ref{lemma:2approx} (the version of
Lemma~\ref{lemma:pi_g} modified for IID service times), we now just
need to modify the proof of Theorem~\ref{thrm:pi_g}.

\proof{Proof of Theorem~\ref{thrm:pi_EDF}} Again, the proof is by
induction the number of jobs with non-zero value that are remaining to
be scheduled. When there is a single job, $\pi^*$ and $\pi^{EDF}$ are
the same and so the bound holds.

Now we need to modify the induction step of
Theorem~\ref{thrm:pi_g}. The Bellman recursion and
Lemma~\ref{lemma:mono_job} give us the following:
\begin{align*}
  V_t^*(s)
  &= \sum_{j \in \Jcal_*} \E[v_j(t + \sigma_j)] + \E[V_t^*(\tilde{S}(s, A_*))]
  \leq \sum_{j \in \Jcal_*} \E[v_j(t + \sigma_j)] + \E[V_t^*(\hat{S}(s, A_*))]
\end{align*}
Let $A_{EDF}$ be the schedule chosen by $\pi^{EDF}$ and let
$\Jcal_{EDF}$ be the set of jobs in
$A_{EDF}$. Lemma~\ref{lemma:2approx} gives us that
$\E[V_t^*(\hat{S}(s, A_*))] = \E[V_t^*(\hat{S}(s, A_{EDF}))]$. Hence,
\begin{align*}
  V_t^*(s)
  &\leq \sum_{j \in \Jcal_*} \E[v_j(t + \sigma_j)] + \E[V_t^*(\hat{S}(s, A_*))]
  = \sum_{j \in \Jcal_*} \E[v_j(t + \sigma_j)] + \E[V_t^*(\hat{S}(s, A_{EDF}))]
\end{align*}
As before, the set of free machines creates a bijection between
$\Jcal_*$ and $\Jcal_{EDF}$. For each $e \in \Jcal_{EDF}$ we can map a
unique $i(e) \in \Jcal_*$. Using this bijection along with
Lemma~\ref{lemma:vm} gives us the following:
\begin{align*}
  V_t^*(s)
  &= \sum_{j \in \Jcal_*} \E[v_j(t + \sigma_j)] + \E[V_t^*(\hat{S}(s, A_{EDF}))]\\
  &\leq \sum_{j \in \Jcal_*} \E[v_j(t + \sigma_j)]
  + \sum_{e \in \Jcal_{EDF}} \E[v_e(t + \sigma_e)]
  + \E[V_t^*(\tilde{S}(s, A_{EDF}))]\\
  &= \sum_{e \in \Jcal_{EDF}}\E\left[v_e(t + \sigma_e) + v_{i(e)}(t + \sigma_{i(e)})\right]
  + \E[V_t^*(\tilde{S}(s, A_{EDF}))]
\end{align*}
Now we use the fact that $\E[v_{i(e)}(t + \sigma_{i(e)})] \leq M$ and the fact that
$\E[v_e(t + \sigma_e)] \geq m$:
\begin{align*}
  V_t^*(s)
  &= \sum_{e \in \Jcal_{EDF}}\E\left[v_e(t + \sigma_e) + v_{i(e)}(t + \sigma_{i(e)})\right]
  + \E[V_t^*(\tilde{S}(s, A_{EDF}))]\\
  &\leq \sum_{e \in \Jcal_{EDF}}\left(\E\left[v_e(t + \sigma_e)\right] + M\right)
  + \E[V_t^*(\tilde{S}(s, A_{EDF}))]\\
  &\leq \sum_{e \in \Jcal_{EDF}}\left(\E\left[v_e(t + \sigma_e)\right] + M\frac{\E[v_e(t + \sigma_e)]}{m}\right)
  + \E[V_t^*(\tilde{S}(s, A_{EDF}))]\\
  &\leq (1 + M/m)\sum_{e \in \Jcal_{EDF}}\E\left[v_e(t + \sigma_e)\right]
  + \E[V_t^*(\tilde{S}(s, A_{EDF}))]
\end{align*}

As in the proofs of Theorem~\ref{thrm:pi_g} and
Theorem~\ref{thrm:pi_G}, we can conclude by applying the induction
hypothesis.

% Now applying the induction hypothesis gives us the result:
% \begin{align}
%   V_t^*(s)
%   %
%   &\leq (1 + M/m)\sum_{e \in \Jcal_{EDF}}\E\left[v_e(t + \sigma_e)\right]
%   + \E[V_t^*(\tilde{S}(s, A_{EDF}))]\\
%   %
%   &= (1 + M/m)\sum_{e \in \Jcal_{EDF}}\E[v_e(t + \sigma_e)]
%   + (1 + M/m)\E[V_t^{EDF}(\tilde{S}(s, A_{EDF}))]\\
%   %
%   &=(1 + M/m)\left(\sum_{e \in \Jcal_{EDF}}\E[v_e(t + \sigma_e) ]
%     + \E[V_t^{EDF}(\tilde{S}(s, A_{EDF}))]\right)\\
%   %
%   &= (1 + M/m)V_t^{EDF}(s)
% \end{align}
\qed
\endproof

To prove Corollary~\ref{cor:pi_EDF}, we simply need to specialize
Theorem~\ref{thrm:pi_EDF} to the case of step-wise decay functions:

\proof{Proof of Corollary~\ref{cor:pi_EDF}} In this case, $\E[v_j(t +
\sigma_j)] = \P(t + \sigma_j \leq d_j) = F(d_j - t)$. Therefore, $M
\leq 1$ and $m \geq p_{min}$.  \qed
\endproof

\section{Proof of Proposition~\ref{prop:ordering}}
Since $x \mapsto \frac{c}{x}$ is a convex function on $x > 0$ for any
$c > 0$, Jensen's inequality gives us that
$$
\E\left[\left.\frac{\sigma_{max}}{\sigma_{min}}\right| \sigma_{max}\right]
\geq \frac{\sigma_{max}}{\E[\sigma_{min} | \sigma_{max}]}.
$$
Since $x \mapsto \min_j x_j$ is concave for $x \in \Rbb^J$, Jensen's
inequality tells us that $\E[\sigma_{min} | \sigma_{max}] \leq \min_j
\E[\sigma_j | \sigma_{max}]$. In addition, $\sigma_{max}$ puts an
upper bound on $\sigma_j$ so $\E[\sigma_j | \sigma_{max}] \leq
\E[\sigma_j]$. Therefore,
$$
\E\left[\left.\frac{\sigma_{max}}{\sigma_{min}}\right| \sigma_{max}\right]
\geq \frac{\sigma_{max}}{\min_j \E[\sigma_j]}.
$$
The law of iterated expectation gives us that
$$
\E\left[\frac{\sigma_{max}}{\sigma_{min}}\right] \geq
\frac{\E[\sigma_{max}]}{\min_j \E[\sigma_j]} = \Delta.
$$
Since $\sigma_{max} \geq \sigma_{min}$, we can conclude that
$$
1 + 2\E\left[\frac{\sigma_{max}}{\sigma_{min}}\right]
\geq 2 + \E\left[\frac{\sigma_{max}}{\sigma_{min}}\right]
\geq 2 + \Delta.
$$

To show that $1 + M/m \geq 2$, we simply note that $M \geq m$ so $M/m \geq 1$.

\section{Discretizing the Lognormal Distribution\label{app:lognormal}}
Given parameters $\ell$, $m$, $s$, and a standard normal random
variable $Z$, $\sigma$ is lognormal if
\begin{equation}
  \sigma = \ell + e^{m + s Z}.
\end{equation}
We will assume that $\sigma$ is measured in minutes.  The cumulative
distribution function of $\sigma$ is
\begin{equation}
  F(x; \ell, m, s) = \left\{
    \begin{array}{lr}
      0 &, x \leq \ell\\
      \frac{1}{2}\left(1 + \erf\left(\frac{\ln(x-\ell) - m}{s\sqrt{2}}\right)\right) &, x > \ell
    \end{array}
  \right.
\end{equation}
where $\erf(\cdot)$ is the error function
\begin{equation}
  \erf(x) = \frac{2}{\sqrt{\pi}}\int_0^x e^{-t^2}dt.
\end{equation}
Assume we use a discretization of $\delta$ with $T$ time slots. Then
we can then take the unnormalized PMF of our discretized lognormal
random variable as
\begin{equation}
  \tilde{p}_{ln}(t; \ell, m, s) = F(t\delta; \ell, m, s) - F((t - 1)\delta; \ell, m, s)
\end{equation}
and the PMF is then given by
\begin{equation}
  p_{ln}(t; \ell, m, s)
  = \frac{\tilde{p}_{ln}(t; \ell, m, s)}{\sum_{t'=1}^T \tilde{p}_{ln}(t'; \ell, m, s)}.
\end{equation}

\section{Randomly Generating Patient Health Decay Functions}
There are several ways of randomly generating continuous piecewise
linear value decay functions. For example, we could take IID samples
from a $\text{Uniform}[0,1]$ distribution, sort the samples in
decreasing order, and then linearly interpolate between these
samples. However, by the law of large numbers, for a large number of
samples, this procedure will yield a function that is roughly
$t~\mapsto~(1 - t/T)$. As a result, because $T = 144$, if we
na\"ively draw $T$ samples and perform this procedure, the value decay
functions will be roughly the same. Because we are trying to model a
mass casualty incident with significant heterogeneity, we need a more
sophisticated method.

To randomly generate a piecewise linear $v(t)$, we first randomly
sample $v(0)$ from $\text{Uniform}[0,1]$. Let $u$ be another IID
sample from $\text{Uniform}[0,1]$. We then define $v(3) =
\max\set{v(0) - (3/T)u, 0}$. We can similarly generate values for
$v(6)$, $v(9)$, etc. and then linearly interpolate between these
points. This will generate a function that decreases continuously and
in a piecewise linear fashion.

We justify this method of randomly generating value decay functions by
qualitatively comparing our results to the results from
\cite{Sacco_Triage_2005}. We randomly generate every third value
rather than every value so that the resulting value decay function is
``smoother.'' A piecewise linear function is essentially intervals of
linear functions that have been ``glued'' together. We notice that the
decay in \cite{Sacco_Triage_2005} has relatively few of these
intervals. The functions that we randomly generate are qualitatively
similar in that they have at most $T/3$ of these
intervals. Furthermore, note that we use 10 minute time
intervals. \cite{Sacco_Triage_2005} used 30 minute intervals (i.e. 3
time units in our model) so our method is in line with the medical
literature.

\end{APPENDICES}

\end{document}